\documentclass[pdftex,nofootinbib,aps,amsmath,amsfonts,amssymb,superscriptaddress,twocolumn]{revtex4-1}
\pdfoutput=1
\usepackage[pdftex]{graphicx}
\usepackage{amsmath}
\usepackage{wasysym}
\usepackage{float}
\usepackage{color}
\usepackage{hhline}
\usepackage{colortbl}
\newcolumntype{L}[1]{>{\raggedright\let\newline\\\arraybackslash\hspace{0pt}}m{#1}}
\newcolumntype{C}[1]{>{\centering\let\newline\\\arraybackslash\hspace{0pt}}m{#1}}
\newcolumntype{R}[1]{>{\raggedleft\let\newline\\\arraybackslash\hspace{0pt}}m{#1}}


\begin{document}

\title{Classical and fluctuation-induced electromagnetic interactions
  in micronscale systems: designer bonding, antibonding, and Casimir
  forces}

\author{Alejandro W. Rodriguez} 
\email{corresponding author: arod@princeton.edu}
\thanks{corresponding author.}
\affiliation{Department of Electrical Engineering, Princeton
  University, Princeton, NJ 08540}
\author{Pui-Chuen Hui}
\affiliation{School of Engineering and Applied Sciences, Harvard
  University, Cambridge, MA 02138}
\author{David N. Woolf}
\affiliation{School of Engineering and Applied Sciences, Harvard
  University, Cambridge, MA 02138} 
\author{Steven G. Johnson}
\affiliation{Department of Mathematics, Massachusetts Institute of
  Technology, Cambridge, MA 02139}
\author{Marko Loncar}
\affiliation{School of Engineering and Applied Sciences, Harvard
  University, Cambridge, MA 02138} 
\author{Federico Capasso} 
\affiliation{School of Engineering and Applied Sciences, Harvard
  University, Cambridge, MA 02138}

\begin{abstract}
  Whether intentionally introduced to exert control over particles and
  macroscopic objects, such as for trapping or cooling, or whether
  arising from the quantum and thermal fluctuations of charges in
  otherwise neutral bodies, leading to unwanted stiction between
  nearby mechanical parts, electromagnetic interactions play a
  fundamental role in many naturally occurring processes and
  technologies.  In this review, we survey recent progress in the
  understanding and experimental observation of optomechanical and
  quantum-fluctuation forces. Although both of these effects arise
  from exchange of electromagnetic momentum, their dramatically
  different origins, involving either real or virtual photons, lead to
  different physical manifestations and design
  principles. Specifically, we describe recent predictions and
  measurements of attractive and repulsive optomechanical forces,
  based on the bonding and antibonding interactions of evanescent
  waves, as well as predictions of modified and even repulsive Casimir
  forces between nanostructured bodies. Finally, we discuss the
  potential impact and interplay of these forces in emerging
  experimental regimes of micromechanical devices.
\end{abstract}



\maketitle

\section{Introduction}
\label{sec:intro}

Light can exert a force.  Although this statement is uncontroversial
today, its confirmation was a remarkable triumph a little over a
century ago\cite{Nichols1903}, only a couple of centuries after
scientists had finally managed to show that light moves at all (as
opposed to appearing instantaneously). Much of the early treatment of
this force was limited to light incident on planar surfaces described
by some absorption and reflection coefficients, in which the resulting
``radiation pressure'' can be explained by intuitive arguments that
continue to dominate pedagogical materials. Perversely, this classical
force is easiest to understand through a quantum
picture~\cite{Jackson98}: since a photon with energy $U$ has
relativistic momentum $U/c$, a black surface that completely absorbs
normal-incident light with power $P$ should experience a force $P/c$,
corresponding to the rate at which it receives momentum.  Similar
arguments apparently imply, from conservation of momentum, that light
can only ``push'' on a flat surface (with a force of at most $2P/c$)
and can never ``pull.''

However, in microstructured systems, the forces induced by
electromagnetic waves are complicated by a number of additional
possibilities. One can consider forces induced by guided waves,
resonant modes, and evanescent waves in addition to forces exerted by
incident waves from vacuum, which can both change the nature of the
force and (thanks to spatiotemporal localization) greatly enhance its
strength. In systems with multiple components, one component can
either pull or push a neighboring component, since
momentum-conservation restrictions only apply to the net force on all
components rather than to the force on any individual object.  Even
for an isolated object, focused beams on a small object can scatter
oblique light forward, a transfer of momentum from the object to the
wave that creates an optical ``tractor beam''~\cite{Ashkin86}.  We
review many of these possibilities and their applications in
Section~\ref{sec:classical}.

In addition to these geometric and localization effects,
Section~\ref{sec:fluct} reviews another consideration that arises for
optical forces at submicron scales: classical optical forces arise
from external sources of electromagnetic fields, but there are also
\emph{internal} sources, namely thermal and quantum charge
oscillations. Most famously, the vibrations of matter give rise to
thermal radiation, the familiar glow of hot objects~\cite{Zhang07},
but the same fields carry momentum as well as energy.  Intuitively,
two hot objects will push each other apart with their thermal
radiation, but this picture is incomplete because it does not include
ambient radiation, and the effect is dramatically altered for objects
in thermal equilibrium with their environment.  For well-separated
objects in thermal equilibrium, the omnidirectional radiation from
both the objects and their environment exactly cancels and there is no
net force on any object, but for surfaces at submicron separations the
evanescent coupling and other effects tend to produce an
\emph{attractive} force.  In the limit of zero temperature, this
attractive force remains due to quantum fluctuations and is known as a
``Casimir'' force~\cite{milonni}.  For a single pair of atoms, the
same phenomenon is a van der Waals force known as the
``London-dispersion'' force' (or the ``Casimir--Polder'' force once
wave effects are included)~\cite{isra}.  (Although such forces have a
sometimes bewildering variety of theoretical descriptions, ranging
from zero-point energy sums to path-integral models, it turns out that
all of these expressions are mathematically equivalent to the forces
of fields produced by vibrating charges in
matter~\cite{Johnson11:review}.) Casimir forces between parallel
metallic surfaces, first predicted in 1948 and reviewed in
Section~\ref{sec:cas-exps}, were finally observed quantitatively in
1978~\cite{Overbeek78} and have subsequently been measured in a wide
variety of microelectromechanical systems (MEMS) with increasing
precision~\cite{Rodriguez11:review}.  However, until the last decade,
both theoretical predictions and experiments were limited to planar or
near-planar geometries with the exception of a handful of theoretical
special cases.  This has now changed, thanks to both powerful new
computational tools and rapidly expanding capabilities of
micromechanical experiments, enabling an explosion in \emph{designer}
Casimir forces (reviewed in Section~\ref{sec:cas-designs}) with exotic
properties extending far beyond simple attraction between parallel
surfaces.  Because Casimir forces remain when all other
electromagnetic interactions (external fields, static charges, and so
on) are removed, Section~\ref{sec:cas-tech} explains that they
represent both an ultimate limitation and, potentially, an opportunity
for MEMS devices as they approach the nanoscale.

\section{Classical optical forces}
\label{sec:classical}

Researchers have long pursued the use of electromagnetic waves to
induce mechanical motion. Kepler was the first to hypothesize that
solar radiation is responsible for the deflection of comet tails away
from the sun. By 1903, Lebedew~\cite{Lebedew01} and Nichols and Hull
\cite{Nichols03} had proved Maxwell's hypothesis that light impinging
on a thin metallic disk in vacuum would induce measurable motion. Over
the course of the next century, applications for harnessing the energy
of light were seen in systems ranging from ``Solar Sails" and
accelerators~\cite{Salamin02} to optical traps and
tweezers~\cite{Ashkin86,Chu91,Grier03,Dholakia05}. In the last decade,
interest in near-field optical interactions has steadily increased, as
on-chip optical circuitry has presented viable alternatives to slower
electronic systems.  The initial single-beam trapping experiment by
Ashkin et. al.~\cite{Ashkin86} was the first to demonstrate the
usefulness of optical gradient forces for the manipulation of
macroscopic objects: a tightly focused laser beam can trap a spherical
dielectric particle in both normal and tangential directions by
balancing the scattering and gradients forces acting on the particle.
Subsequent experiments demonstrated laser cooling and trapping of ions
and neutral atoms~\cite{Chu86, Chu89, Lett89, Phillips84}, leading to
breakthroughs in various aspects of atomic physics, including the
realization of Bose-Einstein condensates~\cite{Anderson95,Davis95},
quantum simulation of arrays of atoms trapped in optical lattices, and
a new field of atom optics~\cite{Mlynek92}.  The physics of
optomechanics has a strong resemblance with Doppler cooling in
atomic/optical physics~\cite{Aspelmeyer10, Kippenberg07,Kippenberg08},
whose groundbreaking development preceded optomechanics by two
decades, except that in optomechanical systems the predicted quantum
nature is manifested in macroscopic objects.  Instead of the atomic
energy levels being dressed due to strong light--atom interactions, in
optomechanics, the photonic resonant states are dressed due to strong
optomechanical interactions. In addition to dynamical effects,
optomechanical forces also lead to nano- and micro-meter mechanical
displacements with milliwatts or smaller incident powers. Similar to
conventional optical trapping, optical forces in optomechanical
systems can be categorized into radiation, gradient, or photothermal
forces. In the following sections, we briefly review the basic physics
of gradient optical forces in micronscale systems and survey recent
developments paving the way toward designable interactions in a wide
range of optomechanical systems, where the amplitude, wavelength, and
phase of incident light can be used to obtain tunable attractive and
repulsive forces.

\subsection{Radiation, gradient, and photothermal forces}
\label{sec:radgradtherm}

\begin{figure*}[tb!]
  \centering
   \includegraphics[scale=0.46]{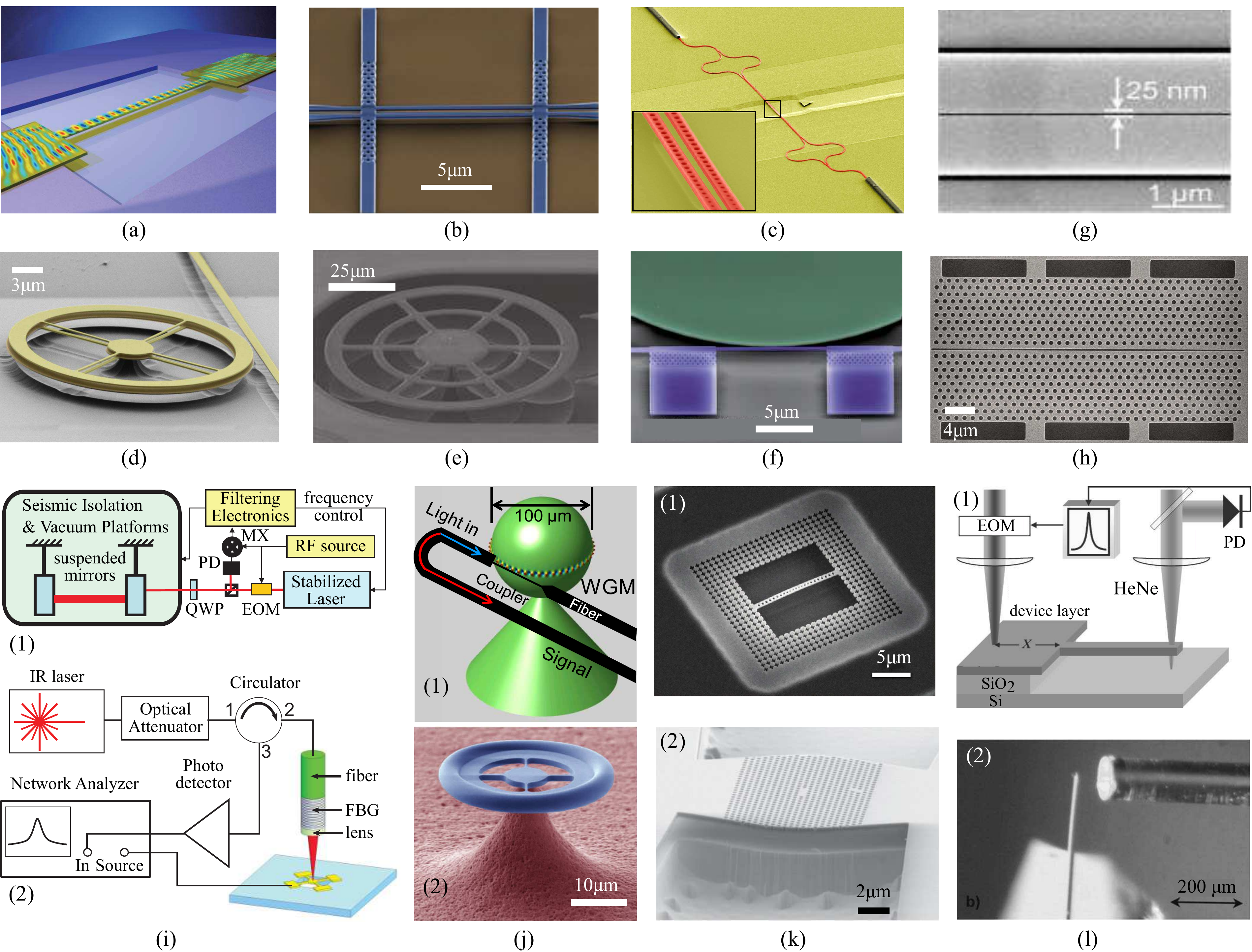}
   \caption{Selected optomechanical structures actuated by radiation,
     gradient, and photothermal forces. Gradient-based designs
     include: (a) A silicon waveguide acting as a doubly-clamped beam
     coupling light incident from the underlying buried oxide layer,
     exhibiting attractive, lateral gradient forces and Duffing
     nonlinearities~\cite{Li08}. (b) Coupled silicon waveguides where
     attractive and repulsive optical forces are exerted by
     controlling the relative phase of incoming light via a
     Mach-Zehnder interferometer~\cite{Li09a}. (c) Coupled silicon PhC
     nanobeam cavities where attractive optical forces are observed in
     the presence of incident incoherent light under atmospheric
     conditions~\cite{Deotare12}. (d) Coupled silicon-nitride ring
     resonators and silica microdisks supporting whispering gallery
     modes leading to attractive and repulsive
     forces~\cite{Wiederhecker09,Rosenberg09,LinRo09}. (f) Dispersive
     and dissipative reactive forces realized in a system of silicon
     waveguides coupled to a silicon microdisk~\cite{Li09b}. (g)
     Gold-coated silicon nitride plasmonic
     waveguides~\cite{Thijssen13} and (h) silicon PhC
     slot-waveguides~\cite{Safavi-Naeini10} exhibiting tightly
     confined electric fields inside the small air gaps.
     Radiation-based designs include: (i-1) Fabry--Perot cavities
     formed by large, kg-scale mirrors suspended by pedulums operating
     at very high powers and exhibiting optical
     instabilities~\cite{Corbitt07}. (i-2) High-finesse Fabry-Perot
     cavities formed by gold-palladium rectangular
     mirrors~\cite{Zaitsev11}. (j-1) Silica microspheres supporting
     optical and acoustical whispering gallery modes where excitation
     of mechanical modes is facilitated by stimulated Brillouin
     scattering~\cite{Tomes09,Bahl11}. (j-2) Silica microtoroids
     supporting whispering gallery modes used to demonstrate coherent
     quantum coupling~\cite{Verhagen12}. (k-1) Silicon optomechanical
     crystals with phononic shields that have been optomechanically
     cooled to the quantum ground state~\cite{Chan11}. (k-2) InP
     photonic crystal cavities designed to have strong localization of
     both optical and mechanical
     modes~\cite{Gavartin11}. Photothermal-based designs include:
     (l-1) Silicon cantilevers actuated by thermal stress induced by
     absorption of visible HeNe light~\cite{Ilic05}.  Here, the
     photothermal effect is modulated by a Fabry-Perot cavity formed
     by the silicon cantilever and the bottom silicon substrate.
     (l-2) Combination of radiation pressure and photothermal effects
     manifested in an optomechanical system comprised of an optical
     fiber above a reflective gold-coated silicon
     cantilever~\cite{Metzger04}.}
     \label{fig:collage}
\end{figure*}

Radiation pressure involves transfer of momentum via propagating waves
to a compliant object. While a photon carrying momentum
$\hbar\omega/c$, or alternatively incident light carrying power $P$,
is bounded by momentum conservation to contribute net pressure $\leq
2P/c$ (assuming 100\% of the light is reflected from the object), the
small momentum imparted by the photon can be drastically enhanced by
introducing an optical cavity. Assuming negligible lossses, the force
on any individual part of a cavity can be many times larger than the
net force on the cavity as the photon continues to exchange momentum
with it over the cavity lifetime $\tau$, thereby enhancing the force
by a factor of $\tau$. The canonical example of such
radiation-pressure enhancements in optomechanical systems is a simple
Fabry-Perot cavity formed by two highly reflective mirrors in which
one of the mirrors is allowed to move, leading to a number of
observable mechanical effects~\cite{Braginsky67}. Demonstrations of
radiation pressure enhancement based on this principle cover a wide
spectrum of length scales and designs, from large-scale mirrors and
Fabry-Perot cavities formed by highly reflective Bragg gratings
[Fig.~\ref{fig:collage}(i)] to micronscale whispering gallery modes
[Fig.~\ref{fig:collage}(j)] of microring resonators~\cite{Ding10},
where light circulates along the circumference of the cavities. The
increasing demand for compact systems with smaller features has led to
designs with increasing complexity and functionalities, including
omnidirectional photonic-crystal (PhC) waveguides operating near band
edges leading to slow group-velocity modes~\cite{Povinelli04:slow},
PhC membranes with ultra-large mechanical lifetimes~\cite{Bui11},
parallel metallic-plate systems operating in the microwave
regime~\cite{Marcet14}, electrostrictive forces arising from
strain-dependent refractive index changes~\cite{Rakich10}, and even
situations involving exotic materials such as left-handed
materials~\cite{Kemp07} and gain media where the force can pull
instead of push objects~\cite{Mizrahi10}. Designs tailored for
applications in optomechanical systems often involve structures with
co-localized optical and mechanical modes [Fig.~\ref{fig:collage}(k)],
where the combination of large optical quality factors $Q$
(dimensionless lifetimes in the range of tens of thousands or above)
and long-lived mechanical modes (with frequencies in the MHz-GHz) have
enabled exciting and novel demonstrations of optical spring effects,
optomechanically induced self-oscillations, on-chip storage and
manipulation of light pulses~\cite{Weis10} and noise~\cite{Zheng13},
on-chip accelerometers~\cite{Krause12}, microfluidic
sensors~\cite{Bahl13}, and cooling of macroscopic
objects~\cite{Vahala03,Rokhsari05,Kippenberg:07,Eichenfield07,JayichSa08,Kippenberg08,Eichenfield09b,Aspelmeyer10}.

Gradient forces involve interactions induced by incident evanescent or
gradient electromagnetic fields. In optomechanical systems, the most
explored of these interactions is the force induced on a microcavity
by the evanescent field of a nearby waveguide or substrate, where
similar to radiation pressure, the forces are also greatly enhanced by
resonances~\cite{Thourhout10}. These include microcavities coupled to
waveguides~\cite{Eichenfield07,LiPernice09}, free-standing slot
waveguides leading to wideband tuning and low-power optical
modulation~\cite{Fong11,Li10,Zheng12}, PhC nanobeam waveguides coupled
to substrates leading to unusual, non-monotonic attractive
forces~\cite{MaPovinelli10}, zipper-like structures leading to
optically controlled mechanical
transparency~\cite{Chan09,Alegre10,Lin2010,Safavi-Naeini10:EIT},
hybrid plasmonic waveguides leading to deep subwavelength confinement
of light~\cite{Yang11,Li13}, and plasmonic
nanobeams~\cite{Halterman05,Thijssen13} or bowtie
antennas~\cite{Kohoutek11}, some of which are illustrated in
Fig.~\ref{fig:collage}. Tunable optomechanical interactions enabled by
large gradient forces~\cite{Groblacher09,Anetsberger09,Thourhout10}
are currently being explored in applications ranging from optical
cooling~\cite{LinRo09}, optical buffers~\cite{Safavi-Naeini10},
re-configurable filters~\cite{Deotare12}, non-interferometric signal
transduction~\cite{Li09:cantilevers,Lin2010}, and optical actuators
and switches~\cite{suh05,Stomeo10,Cai12,Cai13}. The interaction
between coupled resonances through their evanescent field can lead to
even richer phenomena, such as tunable attractive (bonding) and
repulsive (anti-bonding) gradient forces between nearby optomechanical
objects. The ability to generate repulsive forces and to tune the sign
and magnitude of these interactions by adjusting either the wavelength
or phase of incident light is also poised to make an impact in future
photonic switching~\cite{Huang08} and MEMS devices, where they could
be exploited in conjunction with other forces such as Casimir or
electrostatic forces (as described in Section~\ref{sec:cas-tech}).

Finally, light can lead to mechanical deformations through
photothermal interactions, whereby a movable structure absorbs part of
the incident light that is converted to heat.  For instance, thermal
stresses can arise in suspended devices as a result of the difference
in thermal expansion coefficients between the device layer and
underlying supporting substrate\cite{Blocher12}, leading to bending of
the movable structure.  Such photothermal deformations can often be
enhanced by the introduction of a microcavity. Examples illustrated in
Fig.~\ref{fig:collage}(l) include a gold-coated silicon cantilever
excited by a gold-coated optical fiber placed in close proximity,
forming a Fabry--Perot cavity~\cite{Metzger04}, and a cavity formed by
a cantilever and neighboring substrate, excited with a laser frequency
above the silicon bandgap~\cite{Ilic05}.  While the photothermal
effect also allows for optical spring tuning and dynamic back-action
(mediated by delay due to its finite thermal time constant), the
photothermal effect often presents itself as a competing effect in the
demonstration of several optomechanical
devices~\cite{Metzger04,Jourdan2008,Pernice09:thermal,Troger10,
  Zaitsev11,Zaitsev12,Woolf13}.

\subsection{Designer bonding and antibonding forces}
\label{sec:opt-designs}

Among the successes of optomechanics is unprecedented access to
quantum regimes of macroscopic objects, most commonly realized in
singly resonant optomechanical
structures~\cite{Chan11,Verhagen12,Safavi-Naeini12}.  Other endeavors
for expanding the optomechanical toolbox are noteworthy as well. In
the next few sections, we present a number of recent theoretical and
experimental studies which highlight the breadth of designs and
functionalities in coupled, resonant optomechanical systems subject to
bonding and antibonding forces.

\subsubsection{Theory}
\label{sec:opt-theory}

\begin{figure}[tb!]
  \centering \includegraphics[scale=0.4]{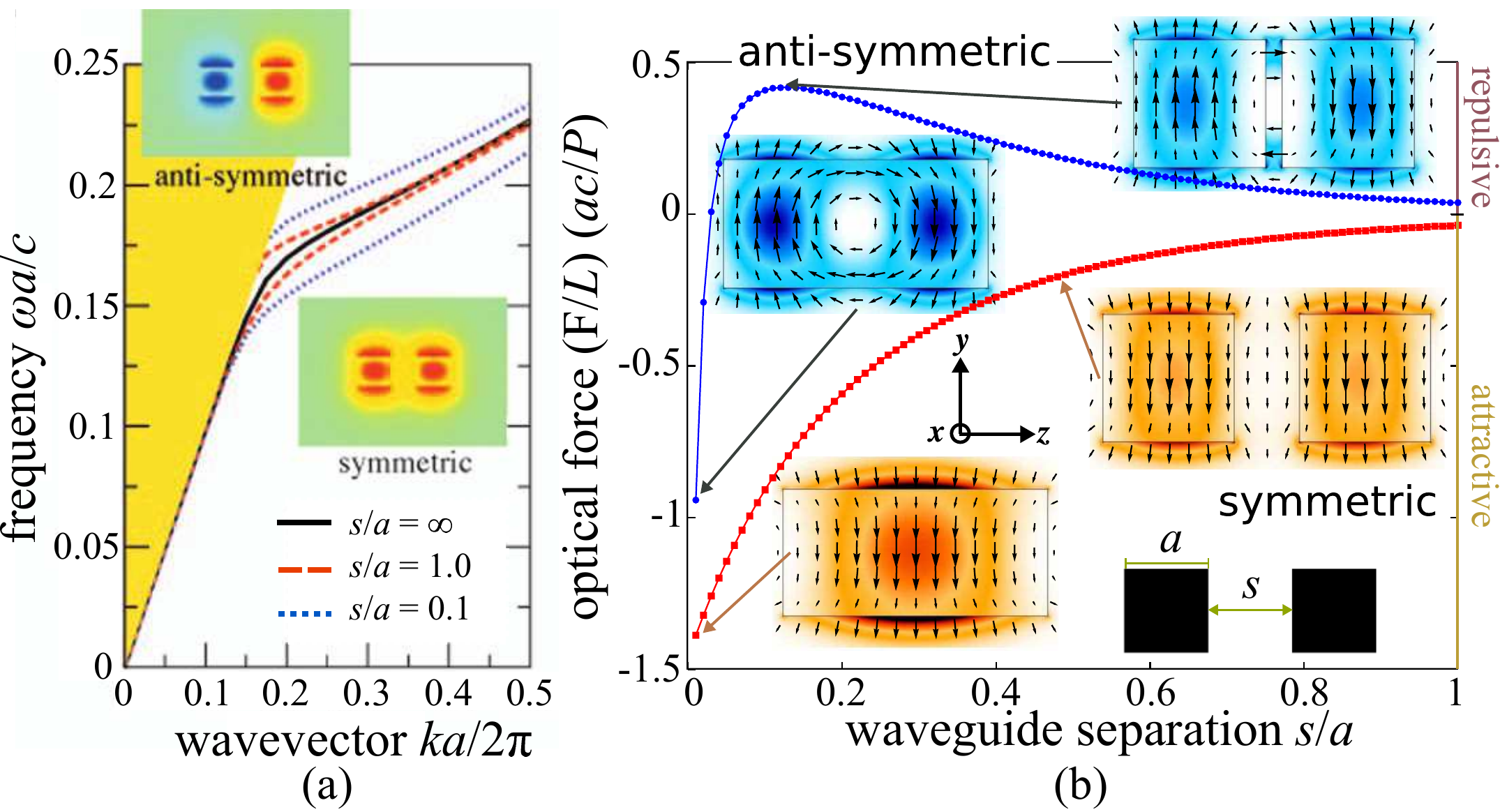}
  \caption{Working principles behind gradient forces in coupled
    optomechanical systems. (a) Dispersion diagram for a system
    involving two identical, co-planar square waveguides at various
    separations $s/a=\infty$ (black solid), 1 (red dashed) and 0.1
    (blue dashed), where $a$ denotes the waveguide
    width~\cite{Povinelli05b}.  Insets show the $E_y$ electric-field
    profile of both bonding (symmetric) and anti-bonding
    (anti-symmetric) modes. As $s$ decreases, the bonding and
    antibonding mode-frequencies experience red and blue shifts,
    respectively.  (b) Normalized gradient force per unit length and
    per incident power for the same system, as a function of $s$ (for
    a fixed incident wavevector), where $+$/$-$ corresponds to
    repulsive/attractive forces~\cite{Oskooi11}. Insets show the
    in-plate electric-field vectorial distribution and total intensity
    of both bonding and antibonding modes. The force amplitude of the
    bonding mode increases monotonically with decreasing $s$, whereas
    the anti-bonding force is only repulsive at separations
    $s/a\gtrsim 0.03$, becoming attractive at short distances.}
   \label{fig:WG_force}
\end{figure}

To illustrate the origin of optical gradient forces between resonant
optomechanical systems we begin by considering one of the first
proposed optomechanical structures exhibiting this effect, involving
two square dielectric waveguides placed in close
proximity~\cite{Povinelli05b}. In such a system, the mutual
interaction of degenerate resonances or guided modes via their
evanescent fields can induce a splitting of the modes into pairs
characterized by attractive and repulsive mechanical forces, analogous
to the well-known bonding and anti-bonding states formed by the level
splitting (avoided crossings) of interacting degenerate states in
quantum systems~\cite{Landau:QM}. The degree of mode splitting from
the initial mode frequencies is controlled by the coupling strength
(or proximity) between the waveguides~\cite{Povinelli05b,Rodriguez11}.
The idea is illustrated in the dispersion diagram of
Fig. \ref{fig:WG_force}(a), which shows changes in the frequency
dispersion of the two waveguide modes as they approach one another
from infinity~\cite{Povinelli05b}. The relationship between changes in
the frequency and mechanical energy of the system can in turn be
understood from a simple heuristic quantum-mechanical
argument~\cite{Povinelli05b}. In particular, assuming that $N$ photons
of frequency $\omega$ and conserved wavevector are coupled into the
waveguide system, the photonic energy $U$ can be written as,
\begin{equation}
  U = N\hbar \omega,
\label{Uopt}
\end{equation}
from which it follows that a small change in the separation will shift
$\omega$ and result in an optomechanical force,
\begin{equation}
  F_\mathrm{om} = -\frac{\partial U}{\partial d} = -\frac{\partial
    (N\hbar\omega)}{\partial d} = -\frac{U}{\omega} \frac{\partial
    \omega}{\partial d}
\label{Fopt}
\end{equation}
acting on both waveguides. As expected, the final expression for the
force does not depend on $\hbar$ but rather on the total stored energy
in the system and can be derived and verified via classical
arguments~\cite{PovinelliJo05,Riboli08,Rakich09}. A useful figure of
merit present in Eq.~\ref{Fopt} is the so-called optomechanical
coupling $g_\mathrm{om} = \frac{\partial \omega}{\partial d}$, which
fascilitates direct comparison of optomechanical forces between
different structures.

Mixing between the modes of the isolated waveguides as they come
together causes a splitting in the frequency, where as expected from a
simple tight-binding picture or from perturbation
theory~\cite{Landau:QM}, they hybridize into symmetric (lower
frequency) and anti-symmetric (higher frequency) modes. Typically,
when the guided waves are in phase (a bonding/symmetric mode), the
electromagnetic energy can be reduced by increasing the field
intensity in high-dielectric regions, leading to an attractive optical
force that pushes the waveguides closer to one another. Conversely,
when the guided waves are out of phase (an anti-bonding/anti-symmetric
mode), the energy can be reduced by increasing the field intensity in
low-dielectric regions, leading to a repulsive force. When mechanical
degrees of freedom are introduced in these coupled structures,
i.e. the waveguides are partially released from the bottom substrate,
these optical forces push the waveguides in an effort to reconfigure
the waveguides and hence lower the total energy. In this particular
example, the polarity of the optical force is predominantly determined
by the relative phases of the electric field in the respective
waveguides, which yields a control parameter to obtain tunable
optomechanical effects, even switching the sign of the force from
attractive to repulsive, an effect that was first observed
experimentally in Refs.~\cite{Li09a,Roels09}. Although the simple
tight-binding picture above is sufficient to explain the main features
of gradient forces at large separations $d \gg \lambda_p$, where
$\lambda_p$ is a characteristic lengthscale corresponding to the
exponential tail of the modes, at shorter separations $d \lesssim
\lambda_p$ it can fail dramatically and one must therefore rely on
exact calculations. These features are illustrated in
Fig.~\ref{fig:WG_force}(b), which shows the forces induced by both the
bonding and antibonding modes of the waveguides over a wide range of
separations. As observed from Fig.~\ref{fig:WG_force}(b), the force
scales exponentially with $d$ at large separations and exhibits a more
complicated $d$-dependence at smaller separations. In this geometry,
non-perturbative effects arising at short separations lead to a
dramatic qualitative change in the behavior of the anti-symmetric
mode, manifested as an increasingly weaker force with decreasing $d$
which ultimately switches sign (becoming attractive) below some
threshold $d \lesssim 0.3a$.

Generalizations of this phenomenon to other coupled-cavity systems
(e.g. microsphere, microdisk, and photonic crystal cavities) have
paved the way for designable gradient forces. While most
optomechanical structures often involve some kind of resonant effect,
technically Eq.~\ref{Fopt} is only applicable in closed systems
comprising lossless (guided) resonances. In the presence of small
losses, e.g. stemming from either radiation or absorption, similar
formulas can be derived (independent of the quantum-mechanical picture
above) which relate forces to the frequencies, incident power, and
lifetimes of the corresponding leaky
modes~\cite{PovinelliJo05,LiPernice09,Rakich09}. In more general
circumstances where there may not be well-defined resonant modes with
negligible loss, or situations where there are superpositions of
resonant modes with other waves (e.g. light from an external source),
modal approaches become problematic. However, because optical forces
are directly related to the solution of scattering problems,
i.e. electromagnetic fields due to incident currents or fields, one
can also frame the calculation of forces using formulations that do
not rely on either mode or energy calculations and which generalize to
other situations of interest, e.g. non-resonant or broad-bandwidth
excitation, and even optical torques~\cite{Grier03}. One such approach
involves computing the force via the integral $\vec{F} = \oint \langle
\vec{T}\rangle \cdot d\vec{S}$ of the time-averaged Maxwell
stress-tensor,
\begin{multline}
  T_{ij} = \frac{1}{2} \Re\left[ \varepsilon_0 \left(E_i E_j
    -\frac{1}{2} \sum_k |E_k|^2\right) \right. \\ \left.+ \mu_0 \left( H_i H_j
    -\frac{1}{2} \sum_k |H_k|^2\right)\right]
\label{eq:Tij}
\end{multline}
around some surface $S$ lying in vacuum~\cite{Jackson98}. Calculation
of stress tensors between vacuum-separated bodies sidestep issues
relating to evaluation of energy densities in lossy
media~\footnote{Evaluation of either energy or stress tensor in
  dissipative media can be problematic~\cite{Jackson98}. However,
  since most cases of interest involve bodies separated by vacuum,
  these issues can generally be ignored.}, and have been performed in
a variety of contexts. For instance, $\langle \vec{T} \rangle$ can be
directly computed via eigenmode calculations in systems with
negligible loss~\cite{Pernice09,WoolfLoncar09}, or more generally by
solving a set of linear equations for the fields in the frequency
domain via finite differences, boundary or finite elements, and
transfer-matrix
methods~\cite{Halterman05,Yannopapas08,XiaoChan08,RohTanabe10,AntonoyiannakisPe10}.
If a broad-band force spectrum is desired, one can also compute stress
tensors via the Fourier transform of a short pulse in the time domain,
yielding the entire spectrum at once~\cite{Rodriguez11}. Modern
numerical methods based on the surface-integral equation formulation
of electromagnetic scattering sidestep altogether the need to
integrate stress tensors over bounding surfaces (which can lead to
numerical problems) or computations of scattered fields, and instead
express the force (or torque) via compact trace formulas that involve
the solution of well-studied linear systems from the boundary-element
method~\cite{Reid14:opt}. Ultimately, since incident light is often
introduced over a narrow range of frequencies, calculations can be
performed expediently and for arbitrarily complicated structures. As
discussed in Section~\ref{sec:cas-theory}, similar numerical
techniques have been developed for computations of fluctuation forces,
where the significantly larger number of radiating centers and
bandwidths complicate matters.

\subsubsection{Recent developments}
\label{sec:opt-recent}

Since the first demonstrations of attractive and repulsive optical
gradient forces between either a silicon waveguide and a silica
substrate~\cite{Li08} or two silicon waveguides~\cite{Roels09}, a
deluge of optomechanical structures actuated by resonantly enhanced
optical gradient forces rapidly emerged. Geometries and materials
explored to attain optomechanical transduction and actuation involving
coupled resonances range from silica and GaAs
microdisks~\cite{Rosenberg09, Jiang09, Ding10}, silicon nitride
microrings~\cite{Wiederhecker09}, and more recently silicon and
InGaAsP photonic crystal (PhC) membranes and
cavities~\cite{Notomi06,TaniyamaNo08,RohTanabe10,Roh10,Deotare12,Gavartin12,Tian2013}.
Each coupled system has its own competitive edge depending on the
desired application, e.g. choice of mechanical modes suitable for
atmospheric operations, choice of materials for thermal power
handling, considerations based on the coupling mechanism or bandwidth
requirements, etcetera, but most designs share the general feature of
frequency-dependent polarity of gradient forces.  Similar to level
repulsion in guided modes of coupled waveguides, resonances with
finite lifetime, be they whispering gallery modes, guided resonances
in one and two-dimensional photonic crystals, or localized modes in
PhC cavities, split into attractive and repulsive force pairs upon
evanescent coupling, enabling actuation of devices by choosing the
corresponding excitation frequencies.

\begin{figure*}[tb!]
  \centering
   \includegraphics[scale=0.41]{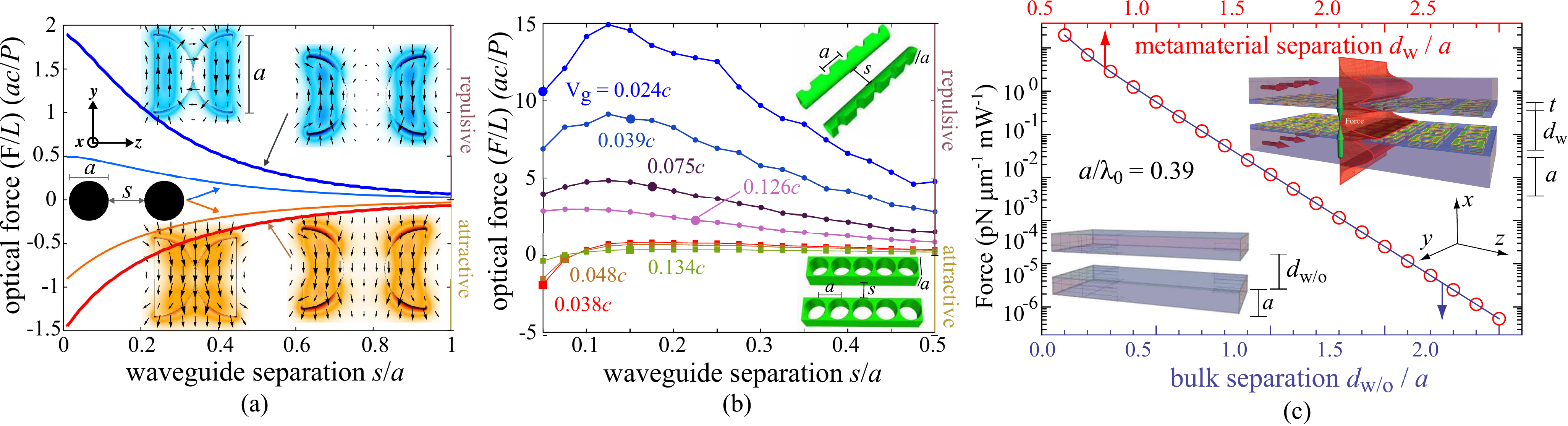}
   \caption{Examples of recent approaches to modifying gradient forces
     between waveguides. (a) Normalized optical force as a function of
     separation $s$ for two waveguide structures consisting of either
     circular (thin lines) or semi-circular (thick lines) cross
     sections~\cite{Favuzzi12}, showing that in contrast to square
     waveguides [Fig. \ref{fig:WG_force}(b)], repulsive forces
     increase monotonically with decreasing $s$. The explanation comes
     from continuity conditions and the fact that convex surfaces tend
     to better concentrate electric fields in low-dielectric regions,
     enabling the frequency of anti-bonding modes to continually
     increase with decreasing $s$. Insets show the in-plane
     electric-field vectorial distribution and total intensity of both
     bonding and antibonding modes. (b) Normalized anti-bonding force
     between coupled PhC waveguides with either square (squares) or
     semi-circular (circles) cross-sections, as a function of $s$. The
     slow-light characteristic of the modes (small group velocity
     $v_g$) enhances the force, an effect that is more prominent for
     semi-circular cross-sections. (c) Slab waveguides cladded with a
     thin metamaterial layer composed of split-ring resonators (top
     inset), designed to reduce the effective distance perceived by
     the evanescent field of interacting modes and leading to force
     enhancements of more than an order of magnitude~\cite{Ginis13}.
     Plot shows the gradient force in the absence (blue line) and
     presence (red circles) of the metamaterial layers, designed such
     that their separation $d_{\mathrm{w}} = d_{\mathrm{w/o}} + 0.5a$,
     where $d_{\mathrm{w/o}}$ and $a$ are the separation and thickness
     of the bulk (unpatterned) slabs.}
   \label{fig:novel_geom}
\end{figure*}

A commonality of current demonstrations of coupled optomechanical
devices is the strategy of tailoring and enhancing optical
interactions.  Since the strength of the optical force is related to
the change in optical energy with respect to mechanical deformations,
techniques of resonantly enhancing light--atom interactions in
photonic chip-based systems were immediately translated to boost the
coupling and force amplitude of optomechanical
systems~\cite{Song05,Zain08,Notomi08,Yinan09,Qimin11}. New directions
of enhancing the transverse attractive and repulsive gradient forces,
apart from the typical approach of employing ultrahigh-$Q$ optical
modes in microspheres or ring
resonators~\cite{PovinelliJo05,NgChan05,Rakich07,Rosenberg09,Jiang09,Wiederhecker10},
have emerged in recent years. Some of these strategies are illustrated
in Fig.~\ref{fig:novel_geom} and include: adopting slow-light Bloch
modes to enhance the intra-cavity optical energy~\cite{Sun12},
designing more complex morphologies to engineer the field
distributions near the interacting
surfaces~\cite{Favuzzi12,Favuzzi13}, and incorporating metamaterials
that locally engineer the dielectric profile experienced by the
evanescent field in the vicinity of the nearby
objects~\cite{Zhang12,Zhang13:meta}, or which effectively reduce the
coupling distance perceived by the fields, thereby ameliorating the
challenge of fabricating free-standing devices with thin sacrificial
layers~\cite{Ginis13}.

Forces between coupled waveguides have been studied in a variety of
planar structures~\cite{Mizrahi07,Riboli08,Pernice09}, including
finite and infinitely thick metal slabs where surface plasmons mediate
the interaction, but whose magnitudes are ultimately limited by high
losses~\cite{WoolfLoncar09}. Spoof plasmons involving sub-wavelength
corrugations on the surface of a semi-infinite metal have also been
explored, demonstrating strong gradient forces at lower frequencies
where losses tend to be smaller~\cite{Woolf14}. However, most of these
systems suffer from limitations similar to those of the original
silicon waveguide geometry of Ref.~\cite{Povinelli05b}, namely the
strength of the repulsive force is bounded. While the attractive force
of a bonding mode monotonically increases in strength as the two
resonant systems approach one other, an antibonding mode is not
guaranteed to generate a repulsive force~\cite{Povinelli05b}. For the
above example of two square waveguides, the relatively large
cross-sectional area and the absence of an air gap between the
structures when touching have been identified as primary reasons for
the transition from repulsion to attraction at short
separations~\cite{Oskooi11}. (The absence of an air gap at short
separations means that the anti-bonding mode cannot continue to
increase its frequency indefinitely~\cite{JoannopoulosJo08-book}.)
Recent work by Oskooi et. al. demonstrated that other waveguide
cross-sections can reverse this trend, and in particular they consider
waveguides with cylindrical and semi-circular cross-sections
[Fig.~\ref{fig:novel_geom}(a)] whose convex inner surface maximize the
presence of air regions as they come together. Even larger forces can
arise in waveguide systems when operating near the band edge of a
guided mode, e.g. induced by introducing a periodic grating along the
invariant direction~\cite{JoannopoulosJo08-book}, due to the smaller
group velocity of these modes. The combination of convex surfaces and
slow-light modes induced by periodicity was shown to lead to orders of
magnitude larger repulsive forces~\cite{Oskooi11}, as illustrated in
Fig.~\ref{fig:novel_geom}(b).

\begin{figure}[tb!]
  \centering \includegraphics[scale=0.445]{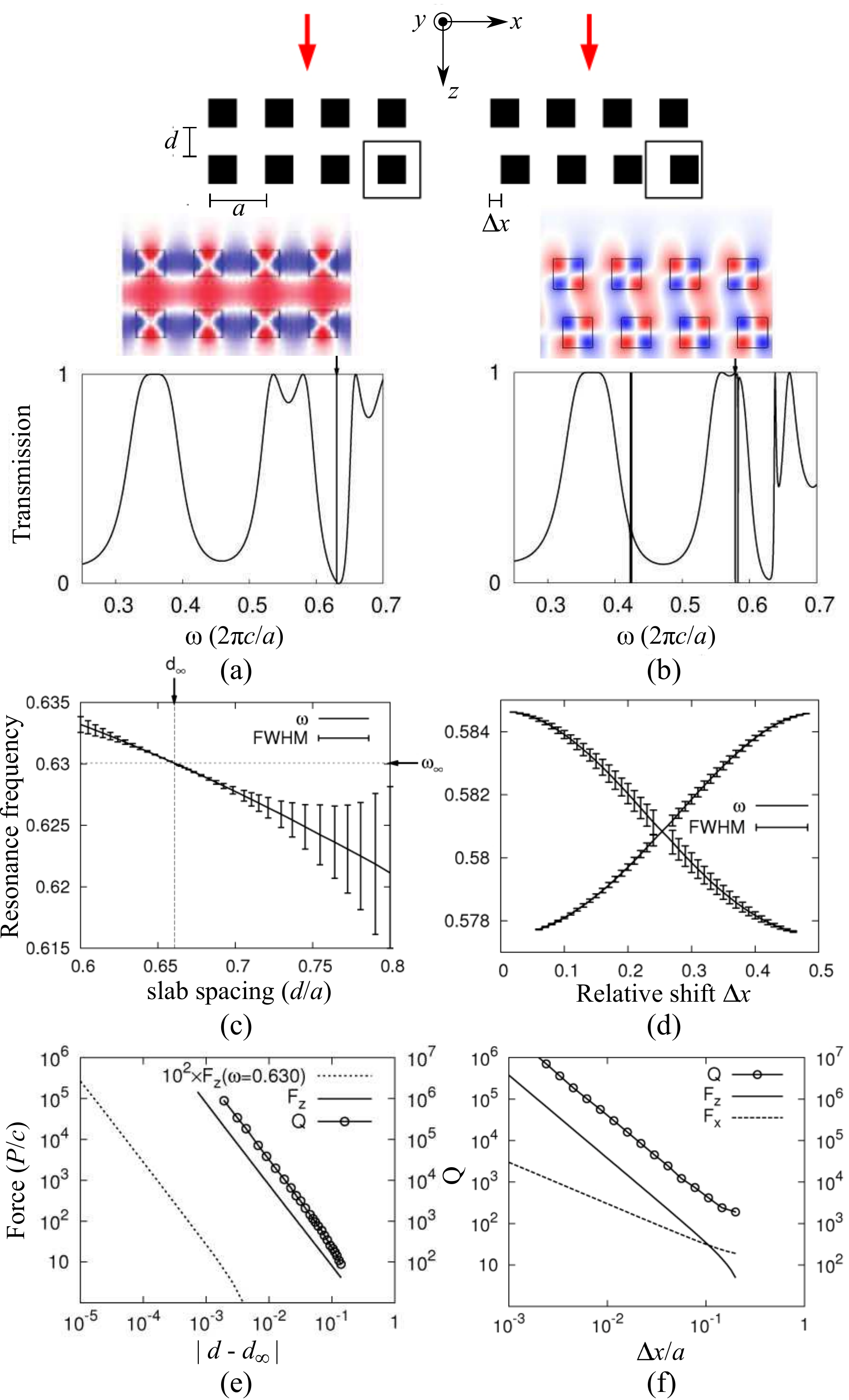}
   \caption{Predictions of ultra-large normal and lateral gradient
     forces in PhC membranes mediated by dark
     modes~\cite{Liu09}. Transmission spectrum of the system for two
     different configurations of membrane separations $d$ and lateral
     translations $\Delta x$, (a) $d=0.65a$, $\Delta x=0$ and (b)
     $d=0.5a$, $\Delta x = 0.15a$, along with the mode profiles of two
     corresponding high-$Q$ dark states appearing at $\omega =
     0.63~(2\pi c/a)$ and $\omega=0.58~(2\pi c/a)$, respectively,
     where $a$ is the membrane period. The dark modes arise due to
     either (a) Fabry--Perot-like interference effects or (b) broken
     symmetry. (c) and (d) show the resonance frequency and linewidth
     of the dark modes in (a) and (b) as a function of $d$ and $\Delta
     x$, respectively. As shown, at $d_\infty$ the membranes do not
     couple to external radiation.  (e) and (f) show the corresponding
     variations in the normal ($F_z$) and lateral ($F_x$) forces, and
     $Q$ for light incident either on (solid line) or slightly detuned
     (dashed line) from resonance (solid line).}
   \label{fig:force_enhance}
\end{figure}

Other recently studied planar-waveguide structures include
metamaterials comprised of metals and dielectrics arranged into
complex microstructures with sub-wavelength
features~\cite{Zhao10,Lapine11,Ginis13,Zhang13:meta,Zhang14}. For
instance, by employing ideas from transformation optics, Ginis
et. al. describe a structure, a thin layer of double-negative
metamaterial involving a double layer of split-ring resonators in
dielectric on top of a dielectric substrate, which allows significant
reduction of the effective optical space between two
waveguides~\cite{Ginis13}. The design and results are presented in
Fig.~\ref{fig:novel_geom}(c--d), which illustrate large enhancements
in the bonding force between the two waveguides. Another class of
artificial structures that show great promise are magnetoelastic or
optomechanical metamaterials, in which gradient forces between movable
elements forming the metamaterial lattice enable on-demand changes in
the metamaterial structure or
lattice~\cite{Lapine11,Zhang13:meta}. Such optomechanical interactions
between ``microscopic'' elements were explored by Lapine et. al. in an
anisotropic magnetic metamaterial structure involving elastic
split-ring resonators, demonstrating strong nonlinear and feedback
effects, e.g. hysteresis, arising from optomechanically induced
changes to the bulk metamaterial properties of the
system~\cite{Lapine11}. These systems present enormous potential for
nonlinear, reconfigurable devices with self-adaptive photonic
functionalities~\cite{Zhang13:meta}. More recently, the effect of
non-locality due to gradient forces mediated by surface-plasmon
polaritons in a wire-based metamaterial medium was
explored~\cite{Zhang14}.

Gradient forces can also be greatly enhanced by planar surfaces
nanostructured at the scale of the incident wavelength, such as those
observed in PhC slabs and microcavities, leading to enhanced gradient
forces over significantly larger areas compared to similar 1d
waveguide structures. For instance, the force between ultra-thin and
high-$Q$ microcavities in PhC membranes was recently shown to lead to
significant optomechanical wavelength and energy
conversion~\cite{Notomi06,TaniyamaNo08,Lu09}. Moreover, the in-plane
periodicity of PhC membrane structures gives rise to leaky
resonances~\cite{JoannopoulosJo08-book} that couple to externally
incident radiation and which can lead to strong bonding and
antibonding forces~\cite{Liu09,RohTanabe10}. Another notable approach
of force enhancement can be found in exploiting symmetry in such
extended periodic structures: by breaking periodicity in PhC gratings
or nanobeams (e.g. by perturbing the alignment of periodic holes),
ultra-large normal and lateral forces can be
engineered~\cite{Liu09,Lu09,RohTanabe10,Favuzzi13}. The idea, first
explored by Liu et. al., relies on the existence of certain leaky
modes, or ``bright'' guided modes of isolated, unstructured slabs that
can couple to external radiation via the periodicity (band folding),
which due to symmetry cannot couple to external
radiation~\cite{Liu09}. While such ``dark states'' have vanishing
bandwidths (infinite lifetimes) in isolated PhC membranes, they can
acquire finite lifetimes (coupling to external radiation) when the two
PhC slabs come into close proximity, as illustrated in
Fig.~\ref{fig:force_enhance}. Specifically, when two such slabs are
evanescently coupled, two interesting phenomena can occur: First, in
the vicinity of certain separations $d_\infty$, the Fabry--Perot-like
interference of bright modes can form high-$Q$ dark states. Second, by
breaking the mirror symmetry of the system, e.g. via a lateral shift
of one of the slabs relative to the other, dark states that otherwise
could not couple to external radiation by symmetry are allowed to. As
shown in Fig.~\ref{fig:force_enhance}, the presence of such ``nearly
dark" modes leads to tunable longitudinal and lateral forces on the
slab. The versatility of actuation and sensitivity of transverse and
lateral optical forces could potentially be employed as an all-optical
three-dimensional accelerometer.

\begin{figure}[th!]
  \centering
   \includegraphics[scale=0.515]{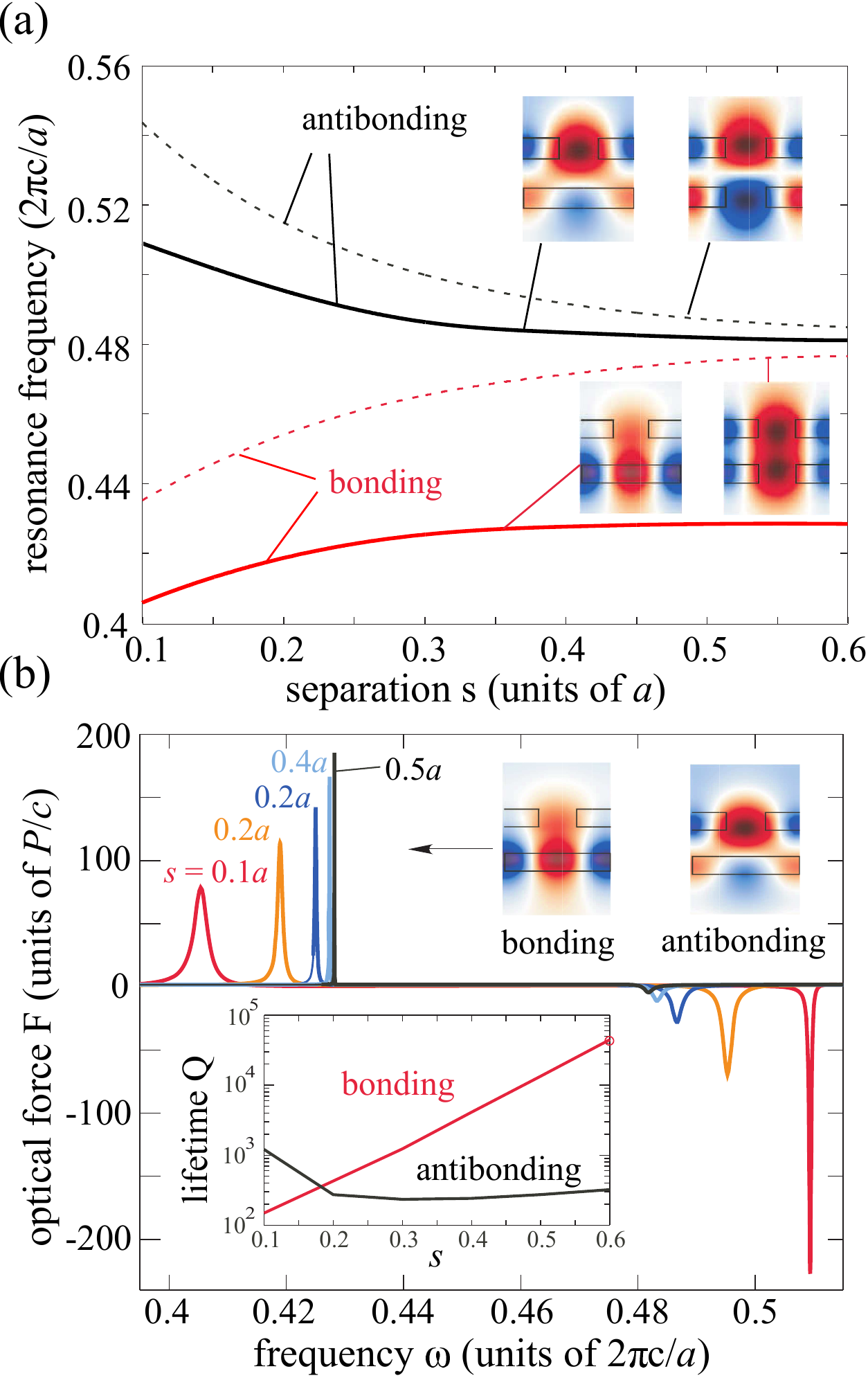}
   \caption{Working principles behind gradient forces in asymmetric
     membrane geometries.  (a) Resonant frequencies (units of $2\pi
     c/a$) of both symmetric PhC--PhC (dashed lines) and asymmetric
     PhC--slab (solid lines) membrane structures of period $a$, as a
     function of membrane separation $s$. Bonding (red) and
     anti-bonding (black) modes are excited by normally incident light
     from above. Insets show the $E_x$ electric-field component for a
     cross-section near resonances at a single $s=0.3a$, delineating
     the in- and out-of-phase characteristics of bonding and
     anti-bonding modes, as well as the asymmetric concentration of
     energy in the PhC--slab structure. (b) Force spectrum of the
     asymmetric structure for multiple PhC-slab separations, obtained
     via the Fourier transform of a short-pulse excitation in
     time~\cite{Rodriguez11}.  Inset shows the corresponding quality
     factor $Q$ as a function of $s$, illustrating the dramatically
     different behavior of bonding versus anti-bonding modes.}
   \label{fig:PhC}
\end{figure}

Attractive and repulsive force pairs also arise in highly asymmetric
planar systems, such as in the system considered by
Ref.~\cite{Rodriguez11} and shown schematically in
Fig. \ref{fig:PhC}(a), consisting of a silicon PhC membrane coupled to
a layered silicon-on-insulator substrate. Here, in contrast to
symmetric PhC membranes, the periodicity of the PhC membrane induces
coupling between the lossy (finite lifetime) leaky resonances of the
membrane and the lossless (infinite lifetime) guided modes of the
silicon-on-silica system. As shown in Fig.~\ref{fig:PhC}(a), level
repulsion arises even when the non-degenerate modes of the two slabs
approach one another (although degenerate modes can also be designed),
leading to bonding and antibonding forces. Two interesting features
stemming from the asymmetry are highlighted here: First, from the mode
profile illustrated in Fig.~\ref{fig:PhC}(b), the bonding mode bears
more resemblance to the slab waveguide mode while the antibonding mode
bears more resemblance to the leaky PhC guided resonance.
Consequently, the bandwidth (lifetime) of the bonding mode has a
significantly stronger dependence on the slab separation
[Fig.~\ref{fig:PhC}(b)], going from $Q=\infty$ at $d=\infty$ to a
finite value $Q\approx 10^2$ at shorter separations.  Second, as a
result of the strong coupling-dependent variation in $Q$ of the
bonding mode, while the repulsive force (antibonding mode) increases
in strength as the membrane separation decreases, the attractive force
(bonding mode) amplitude decreases, as shown in Fig.~\ref{fig:PhC}(b),
in contrast to what is normally observed in most mirror-symmetric
systems. A related structure, involving a nanostructured membrane on
top of a substrate, was recently studied in the metamaterial limit
where the membrane consisted of a plasmonic
metamaterial~\cite{Zhang12}. 

\begin{figure}[t!]
  \centering  
   \includegraphics[scale=0.31]{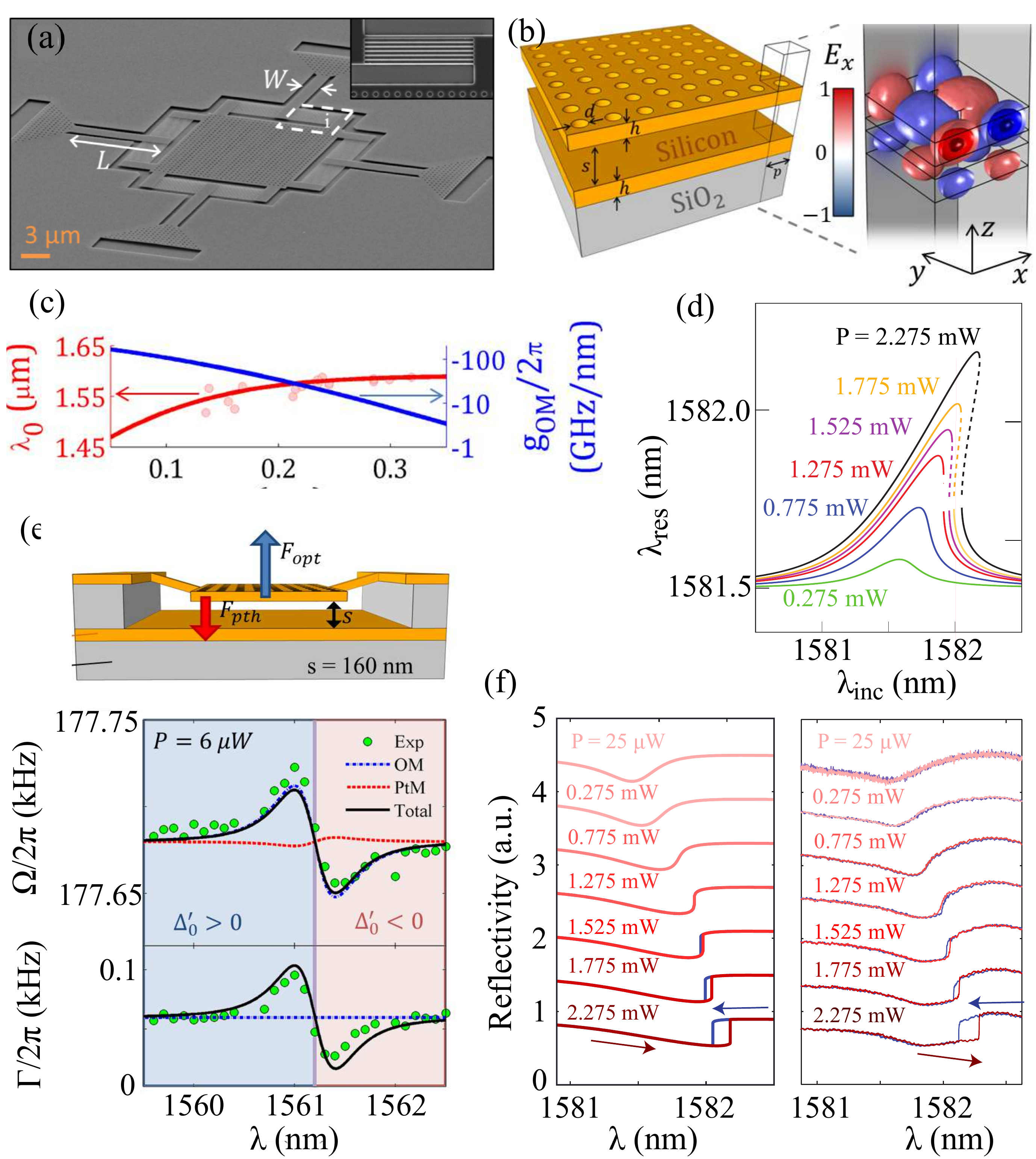}
   \caption{Experimental demonstration of dark-mode repulsive forces
     in PhC membranes~\cite{Hui13}. (a) Electron micrograph of device
     consisting of a silicon PhC membrane suspended above a
     silicon-on-insulator substrate, shown schematically in (b). The
     membrane is a $27.6\mu\mathrm{m}\times 27.6\mu\mathrm{m}$ slab of
     thickness $h=185$nm perforated by a 30$\times$30 array of holes
     with diameter $d=0.414\mu\mathrm{m}$ and period
     $p=0.92\mu\mathrm{m}$. Also show is the $E_x$ electric field
     profile of an anti-bonding dark mode at the incident wavelength
     $\lambda=1581.55$nm. (c) Calculated (solid lines) and measured
     (pink circle) resonance wavelength (red) and optomechanical
     coupling $g_\mathrm{om}$ (blue) of the dark mode as a function of
     the PhC--slab separation $s$. (d) Resonance wavelength
     $\lambda_\mathrm{res}$ as a function of incident laser power $P$
     and wavelength $\lambda_\mathrm{inc}$, for six values of $P$. At
     a certain critical power $P_\mathrm{bis} \approx 1.275$mW, the
     system exhibits bistability stemming from optomechanical and
     photothermal effects, displaying hysteresis beyond a threshold
     power $P_\mathrm{hys} \approx 1.525$mW as the laser wavelength is
     continuously swept backwards (blue line) and forwards (red line),
     as evidenced by the theoretical (1) and experimental (2)
     reflection spectra in (f). (e) Schematic illustration of the
     competing optomechanical and photothermal forces on the membrane,
     along with the measured mechanical frequency $\Omega_m$ and decay
     rate $\Gamma_m$ of the fundamental mechanical mode across the
     optical resonance at incident power $P=6\mu$W. Experimental data
     is fitted to predictions from theoretical models based on the
     coupled-mode theory framework~\cite{Hui13}.}
   \label{fig:dSOI}
\end{figure}

A recent experimental demonstration of repulsive forces that captures
many of the design principles discussed above involves the asymmetric
structure shown in Fig.~\ref{fig:dSOI}(a--b), featuring a tethered
silicon PhC membrane of size $30\mu\mathrm{m} \times 30\mu\mathrm{m}$
suspended above a typical silicon-on-insulator (SOI)
substrate~\cite{Woolf13,Hui13}. Specifically, the optomechanical
properties of the structure were probed and actuated with a band-edge
dark resonance, whose in-plane symmetry and field (phase) profile are
illustrated in Fig.~\ref{fig:dSOI}(b), leading to a repulsive force in
the telecom range. In contrast to approaches based on mirror-symmetric
bodies, experimental access to the dark mode in the asymmetric system
was made possible due to the finite size effect of the membrane as
well as fabrication-induced inhomogeneities, resulting in
significantly lower lifetimes $Q \approx 4400$ than some of the other
recently studied structures, featured in Table~\ref{table:dev}.  For
vertically coupled optomechanical devices like those shown in
Fig.~\ref{fig:collage}(d--e), tunability of the optomechanical
coupling strength is not well controlled or would otherwise require a
new substrate with different sacrificial layer thicknesses.  As shown
in Fig.~\ref{fig:dSOI}(d), wide-range tuning of the optomechanical
coupling strength in the range $g_\mathrm{om}=-2\pi \times
(5,66)$~GHz$/$nm was achieved on the same substrate by engineering the
in-plane compressive stress of the silicon device layer and the
stress-gradient-induced torque, enabling control of the separation
between the membrane and substrate in the range of 120--300nm. In
addition to exhibiting both optical spring effects and dynamic
back-action [Fig.~\ref{fig:dSOI}(e)], the system was shown to exhibit
blue-detuned cooling and red-detuned amplification, in contrast to
what is normally observed in conventional optomechanical systems. Such
unusual effects are a manifestation of the interplay between
photothermal and optomechanical forces. Moreover, optical bistability
is observed as a result of both optomechanical dispersion and
thermo-optic effects [Fig. \ref{fig:dSOI}(f)], occurring whenever the
membrane transitions from one mechanical equilibrium to another
created by the optical potential.

\begin{table*}[t!]
\centering
\begin{tabular}{|C{2.25cm}|C{4.45cm}|C{1.85cm}|C{1.85cm}|C{1.2cm}|C{1.65cm}|C{1.6cm}|C{1.75cm}|}
\hline
\textbf{Team} & \textbf{Structure -- separation (nm)} & \textbf{Mechanical frequency $f_\mathrm{m}$} (MHz) & \textbf{Mechanical quality factor $Q_\mathrm{m}$} & \textbf{Optical $Q_\mathrm{opt}$} & \textbf{Coupling $g_\mathrm{om}/2\pi$ (GHz$/$nm)} & \textbf{OM force (nN$/$mW)} & \textbf{Amplitude (nm$/$mW)} \\ \hline
Eichenfield et. al. 2007~\cite{Eichenfield07} & SiN disk resonator with tapered fiber -- 702nm & $1.93\times 10^{-6}$ & - & $1.1\times10^6$ & - & -0.02 & -324 \\ \hline
Li et. al. 2008~\cite{Li08} & Si waveguide with SiO2 substrate -- 360nm & 8.87 & 1850 & - & - & 0.005 & 2 \\ \hline
Li et. al. 2009~\cite{Li09a} & Laterally coupled Si waveguides -- 100nm & 17.05 & 5300 & - & - & -2.2 (1.1) & 9.6 \\ \hline
Rosenberg et. al. 2009~\cite{Rosenberg09} & Vertically coupled SiO2 ring resonator -- 138nm & 8.3 & 3.95 (in air) & $1.8\times10^{6}$ & 33 & -244 & - \\ \hline
Wiederhecker et. al. 2009~\cite{Wiederhecker09} & Vertically coupled SiN ring resonator -- 640nm & 0.6 & 2 (in air) & $6.8\times10^{4}$ ($2.1\times10^4$) & 1.4 (2) & - & -20 (1) \\ \hline
Eichenfield et. al. 2009~\cite{Eichenfield09b} & Laterally coupled SiN zipper nanobeam cavity -- 120nm & 8 & 11600 (50) (in air) & $3\times10^5$ & 123 & - & - \\ \hline
Roh et. al. 2010~\cite{Roh10} & Bilayer InP PhC membranes -- 200nm & 1.8 & 2 (in air) & 700 (1600) & (44) & (-0.83) & (-0.26) \\ \hline
Deotare et. al. 2012~\cite{Deotare12} & Laterally coupled Si PhC nanobeam cavities -- 70nm & 8 & 17 (in air) & 15000 & 96 &  -1400 & 0.025 \\ \hline
Woolf et. al. 2012~\cite{Woolf13} & Si PhC membrane coupled with an SOI substrate -- 160nm & 0.16 & 2000 & 3400 & -66 & 1.2 & 1 \\ \hline
\end{tabular}
\caption{Representative optomechanical devices based on gradient
  forces along with several corresponding figures of merit, including
  typical mechanical separations, frequencies $f_\mathrm{m}$ and
  quality factors $Q_\mathrm{m}$, measured optical quality factors
  $Q_\mathrm{opt}$, optomechanical couplings $g_\mathrm{om}$ ($+$/$-$
  denotes bonding/anti-bonding modes), optomechanical force
  amplitudes, and static/dynamic mechanical amplitudes. Entries marked
  by hyphens ``-'' denote missing data.  Unless otherwise noted,
  measurements are assumed to be in vacuum.}
\label{table:dev}
\end{table*}

Generally, photonic structures that support many resonances,
e.g. microdisks, PhC slabs, and microspheres, will exhibit complicated
force spectra due to the coupling-induced frequency-splitting among
all modes, especially in the higher frequency range where the density
of states increases. In combination with the complex interactions
between lossless and leaky resonances arising in systems such as the
asymmetric membranes above, these features shed light on new
strategies for engineering resonantly enhanced gradient forces. For
instance, it is possible to perform multi-modal excitations to
engineer the net optical force exerted on the optomechanical structure
and thereby tune the mechanical spring constant dramatically without
perturbing the initial mechanical
equilibrium~\cite{Rodriguez11}. Simultaneous frequency excitations in
systems with sub-micron dimensions can also create highly localized
optical traps~\cite{Rakich07,Rodriguez11}, opening up new avenues of
optomechanical trapping and even exposing the system's mechanical
nonlinearity as the linear response is suppressed.

To summarize the various approaches of gradient force enhancement,
Table~\ref{table:dev} highlights some of the recently studied and
representative gradient-force-actuated optomechanical devices,
comparing their fundamental mechanical frequencies and quality
factors, optical lifetimes, optomechanical coupling strength, and
maximum force/mechanical amplitudes.  One observation is that various
material systems are chosen for considerations such as photothermal
effect, free carrier excitation and stress management.  Another
feature is the successful demonstration of atmospheric operations of
these optomechanical devices which circumvent the need for hermetic
vacuum packaging useful in sensing applications. Finally, we note that
the realization of optomechanical structures with greater design
complexities demands more sophisticated fabrication techniques.  We
anticipate that micro-fabrication advancements, including the ongoing
development and experimental demonstrations of metamaterials in the
micron-scale and multi-layer thin-film assembly assisted by soft
lithography \cite{Gao2014,Soukoulis2011,Zhao2012,Maas2013,Kusaka2014},
will continue to pave the way toward new demonstrations of pronounced
force enhancements with optimized optical designs.

\subsection{Technological Impact}
\label{sec:opt-tech}

\begin{figure*}[th!]
  \centering \includegraphics[scale=0.255]{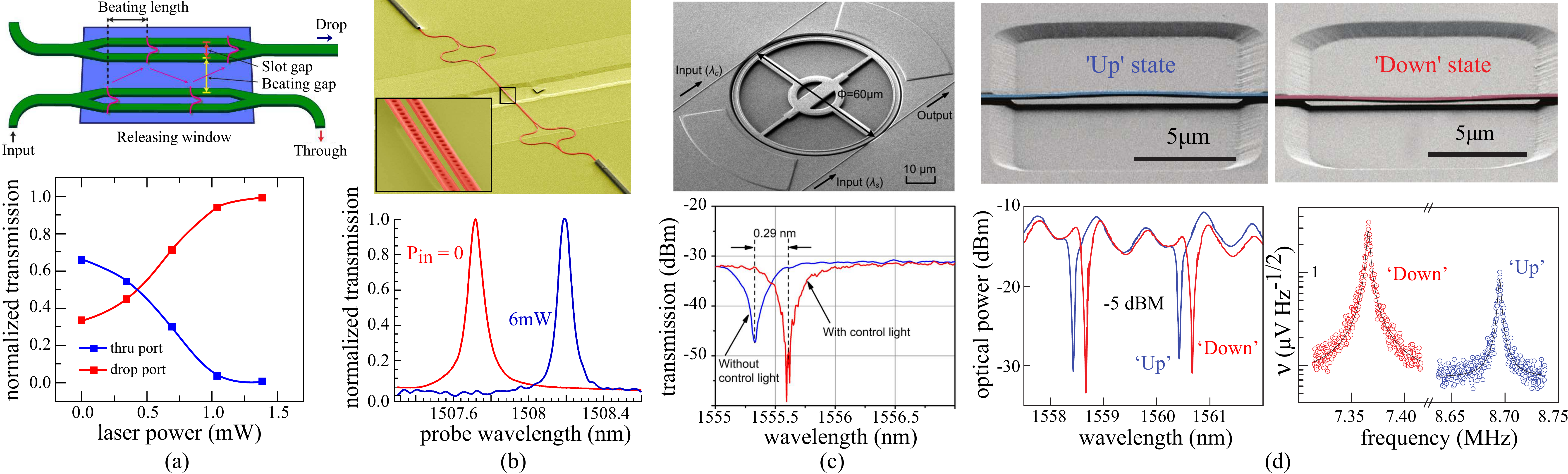}
  \caption{Selected applications of experimentally realized
    optomechanical systems exploiting gradient forces. (a) Optical
    waveguide couplers where on-chip tuning is achieved by strong
    optomechanical coupling in silicon slot
    waveguides~\cite{Fong11}. Plot shows static tuning of the through
    (blue) and drop (red) port transmissions as a function of incident
    laser power at two different probe wavelengths. (b) Coupled
    photonic crystal nanobeam cavities employed as an all-optical
    tunable filter that can be actuated with incoherent light
    sources~\cite{Deotare12}. Plot shows tuning of the cavity
    resonance via thermo-optic, free-carrier, and optomechanical
    effects induced by incident power at 0 (red) and 6mW (blue). (c)
    Optomechanical switching with timescales on the order of tens of
    nanoseconds, where resonance tuning is achieved by controlling the
    coupling between a silicon resonator and an underlying buried
    oxide layer~\cite{Cai13}. Plot shows a zoom-in view of the
    wavelength shift induced by the control light in either the off or
    on state. (d) Optomechanical memory involving bistable transitions
    induced by gradient forces~\cite{Bagheri11}. Plots show the
    transmission (optical power) and thermomechanical ($\nu_n$) noise
    spectra at low input powers when the resonator (shown
    schematically) is in the up (blue) or down (red) states. The 'up'
    and 'down' states enable the possibility of writing nanomechanical
    memory sequences.}
   \label{fig:app}
\end{figure*}

The conspicuous expression of the optical force in nanophotonic
devices facilitates new strategies for achieving reconfigurable and
programmable optical
devices~\cite{FrankPa10,Deotare12,Fong11,Rosenberg09,Cai13}, further
advancing the nascent field of nano-optoelectro-mechanical systems
(NOEMS).  Among coupled optomechanical structures, recently
demonstrated reconfigurable passive optical elements include couplers
with a pair of coupled optomechanical slot waveguides~\cite{Fong11},
broadband all-optical filters that can be controlled by incoherent
light~\cite{Deotare12}, and optical switches based on either coupled
microdisks~\cite{Rosenberg09} or waveguide--substrate
platforms~\cite{Cai13} exhibiting switching times on the order of
nanoseconds. Thus far, despite the relatively large degree of
actuation achieved by optical forces in nanophotonic devices,
electrostatic actuation still outperforms optomechanics in the extent
of actuation~\cite{FrankPa10,Winger11,Poot14}.  However in
applications where operating environments are adverse to systems with
metallization, an all-optical platform could still be desirable.


A subtle effect that the optomechanics community has dealt with in
past years is buckling of the optomechanical devices caused by
compressive stress in device layers, e.g. in typical
silicon-on-insulator substrates~\cite{Sun2012,Iwase2012}, gallium
arsenide~\cite{Midolo2012}, and diamond~\cite{Rath2013}. Such
compressive stresses cause deviations of the fabricated structures
from the desired geometry.  On the one hand, solutions to this problem
include resorting to material systems with \emph{tensile} stress
(e.g. silicon nitride)~\cite{Fong2010,Wiederhecker09}, depositing a
thin layer of materials to compensate the stress~\cite{Hadzialic2010},
or developing structural stress-relief
techniques~\cite{Iwase2012,Krause12}.  On the other hand, some recent
proposals take advantage of the presence of buckling to create
mechanical, bistable states for switching and sensing which basically
eliminate the consumption of holding power.  Examples include
demonstrations of mechanical memory by Bagheri et
al.~\cite{Bagheri11}, optical shock sensors~\cite{Dong14}, and optical
switches~\cite{Intaraprasonk11}. The strong optomechanical strength of
modern designs also offers high readout sensitivity of mechanical
motion, even under atmospheric conditions where mechanical signals
could still be detected above the noise floor in the presence of
strong viscous damping.  Some of the recent demonstration exploiting
readout sensitivity hard to reach by conventional electromechanical
schemes include broad-bandwidth accelerometry in optomechanical slot
waveguides~\cite{Krause12}, particle detection with self-oscillating
toroidal resonators in air~\cite{Liu13a}, optical
switches~\cite{Cai13}, optomechanical AFMs~\cite{LiuKartik12, Miao12},
and microfluidic optomechanical sensing in liquid
environments~\cite{Bahl13, Kim13}. There is also strong drive to seek
applications of optomechanics in the classical and quantum regimes,
particularly in the radio-frequency window.  However, the design of
optomechanical devices that operate in the GHz range and exhibit
efficient optomechanical transduction is a highly non-trivial problem.
For instance, there are significant endeavors to design structures
that feature strong colocalization of photonic and phononic modes that
maximize dispersive coupling~\cite{Oudich2014,El2013,Hsiao2012} and
which enable wavelength conversion (from telecom to telecom and
telecom to microwaves) mediated by GHz mechanical modes in the
sideband-resolved regime~\cite{Safavi-Naeini11, Tian10, Notomi06,
  Hill12, Davanco12, Davanco14, LiuKartik13, Bochmann13}.
Self-oscillating optomechanical oscillators for timing applications in
the radio-frequency window are also explored with phase-noise
suppression schemes incorporated to rival existing technology of
crystal oscillators~\cite{Zheng12, Zheng13, Rokhsari06, Hossein06,
  Hossein08a, Hossein08b, Hossein10, Liu13b, Liu14}. Finally, the
miniaturization of devices based on optomechanical forces will
necessarily lead to other important considerations stemming from other
competing effects, including electrostatic and fluctuation forces, the
subject of the next sections.

\section{Fluctuation forces}
\label{sec:fluct}



As micromechanical devices enter the sub-micron regime,
fluctuation-induced electromagnetic forces such as van der Waals or
Casimir forces become increasingly important, leading for example to
unwanted ``stiction'' between moving
parts~\cite{Serry98,Buks01:mems,Capasso07:review,Genet08}. Unlike
their classical analogue, these interactions have their origins in the
quantum and thermal fluctuation of charges in
bodies~\cite{casimir,lifshitz1,Dzyaloshinskii61,Plunien86,Spruch96,Kardar99,milton01,milton04,milonni,bordag01,Lambrecht02:world,Lamoreaux05,Onofrio06,Lamoreaux07:phystoday,Ball07,Capasso07:review,Buhmann07,Rodriguez11:review,Dalvit11:review,Buhmann12}
and hence persist even in the absence of external inputs. The
volumetric and broadband character of these fluctuations and their
usually small nature makes design, calculations, and measurements of
these forces significantly more challenging than their classical
counterparts, although they can reach atmospheric pressures at
nanometric separations. Theoretical calculations and experimental
measurements were until recently limited to planar or nearly-planar
structures, where forces are usually attractive and monotonically
increasing with decreasing separation. Recent theoretical and
experimental progress, however, is making it possible to study the
ways in which geometry and materials affect both the sign and
magnitude of these interactions, paving the way for potentially new
design principles to enter micromechanical and microfluidic
devices. In the following sections we review the basic physics of
Casimir forces, describe the similarities and differences with
classical optical forces and discuss recent design principles and
experiments that probe the strong material and geometry dependence of
the force, including predictions of repulsive or unusual interactions
between microstructured surfaces. Finally, we discuss recent progress
toward measuring and designing Casimir forces in integrated and
optomechanical devices in which classical forces are used to actuate
as well as combat stiction. For other recent reviews, the reader is
encouraged to look in
Refs.~\cite{Lamoreaux07:phystoday,Ball07,Capasso07:review,Buhmann07,Rodriguez11:review,Dalvit11:review,ReidRo12:review,Buhmann12}

\subsection{Casimir forces}
\label{sec:cas-theory}

The early studies of EM fluctuation forces dates back to the
pioneering work of Johannes D. van der Waals, who predicted an
attractive force between neutral atoms or particles stemming from
quantum dipolar
fluctuations~\cite{London37,ninham,isra,Parsegian06}. Such an
interaction arises whenever a particle acquires a spontaneous dipole
moment, either from quantum or finite-temperature fluctuations,
producing a field that can polarize a nearby particle and lead to a
corresponding dipole--dipole interaction~\cite{Buhmann07}. The
cumulative effect of such dipolar interactions over all frequencies
are so-called ``dispersion'' forces that depend on both the separation
and frequency-dependent polarizabilities of the particles, and which
have come to be known as van der Waals forces in the near field
(interactions dominated by evanescent fields) or Casimir--Polder
forces in the far field (interactions dominated by radiative fields
and retardation effects)~\cite{Casimir48:polder}. Casimir forces are
generalizations of this phenomenon to systems composed of many
particles, such as macroscopic media~\cite{casimir}, in which case the
single-particle picture fails to capture important multiple
scattering, boundary, and wave effects that can lead to strong
modifications of the force laws.

\begin{figure}[tb!]
  \centering
  \includegraphics[scale=0.39]{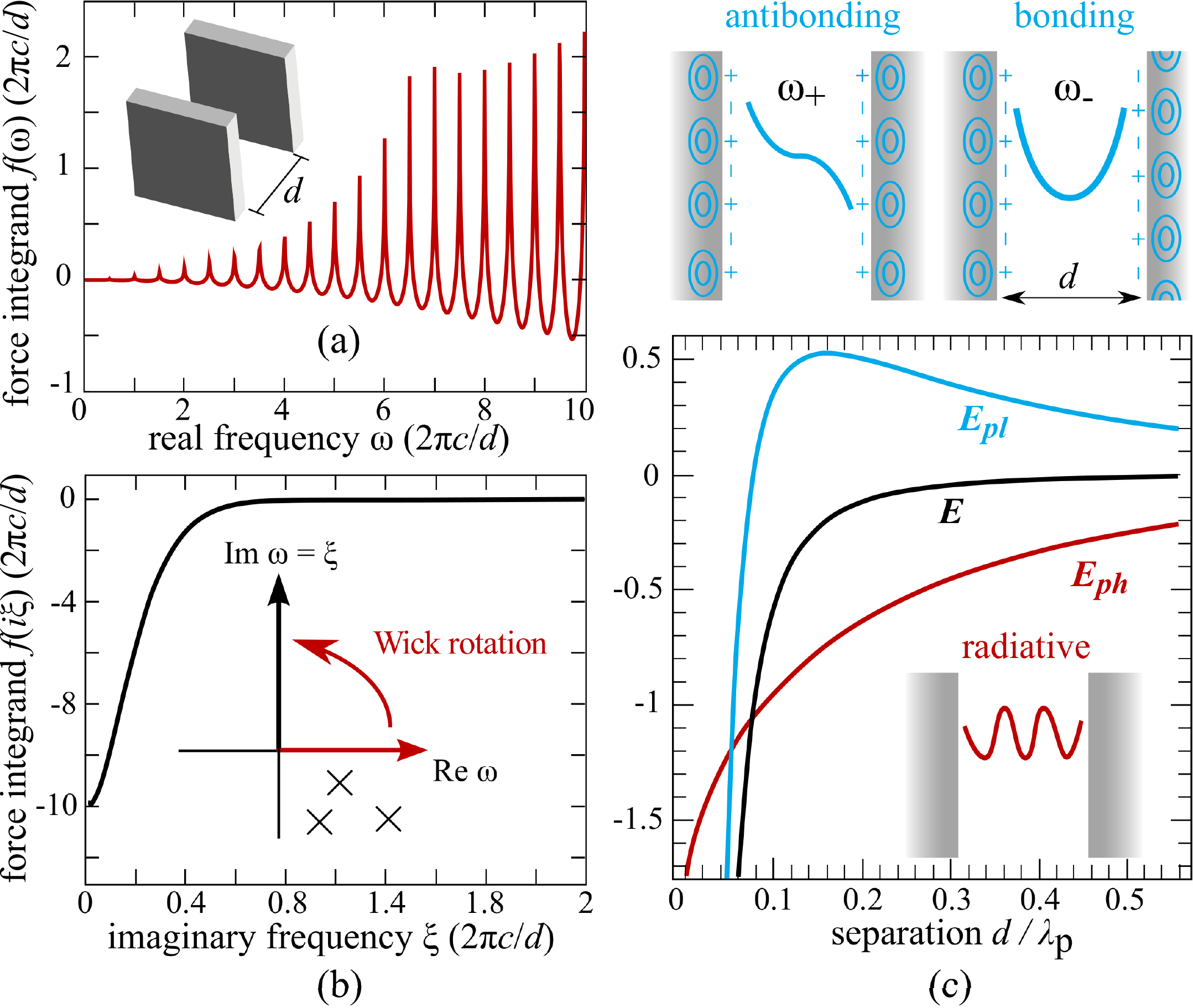} 
\caption{Spectral decomposition of the Casimir force between gold
  plates illustrating the advantages of going into the
  imaginary-frequency axis, as well as the overall impact of
  propagating and evanescent modes on the force. Casimir force
  integrand as a function of (a) real $\omega$ and (b) imaginary
  $\omega\to i\xi$ frequencies, illustrating the transition from a
  broad-bandwidth oscillatory integrand to a smooth, exponentially
  decaying one (adapted from Ref.~\cite{ReidRo12:review}). (c) Casimir
  energy $E$ (black line) decomposed into contributions from
  evanescent/plasmonic $E_{pl}$ (blue line) and propagating/photonic
  $E_{ph}$ (red line) modes, as a function of the plate--plate
  separation $d$ normalized by the plasma wavelength $\lambda_p$. The
  plasmonic energy switches sign from attractive in the near field ($d
  \ll \lambda_p$) to repulsive in the far field ($d \gg \lambda_p$),
  peaking at $d\approx 0.16\lambda_p$, while the radiative energy
  leads to attraction at all
  distances~\cite{Intravaia05}.}  \label{fig:casimag}
\end{figure}

In such situations, a different approach is needed, such as the
framework introduced by Hendrik Casimir in 1948. In particular, rather
than summing the energy contribution of dipoles within the macroscopic
bodies, Casimir showed that it is equivalent to instead consider the
energy stored in the corresponding (induced) electromagnetic
fields. Within this formulation, the quantum-mechanical zero-point
energy of the electromagnetic field can be related to the sum over
modes of the system,
\begin{equation}
  U = \sum_\omega \frac{1}{2} \hbar \omega,
\label{eq:Ucas}
\end{equation}
where $\hbar$ is the reduced Planck's constant and $\hbar\omega/2$ is
the energy of a virtual photon with frequency $\omega$. Application of
Eq.~\ref{eq:Ucas} to Casimir's simplified model of two perfect
electric-conductor (PEC) plates separated by a distance $d$, leads to
the well-known formula for the pressure $P_\mathrm{C} =
-\frac{\partial U}{\partial d}$ between the
plates~\cite{casimir,milonni},
\begin{equation}
  P_\mathrm{C} = -\frac{\hbar c}{240 \pi d^4} = 0.013\frac{1}{d^4}
  \mathrm{dyn}(\mu\mathrm{m})^4 \mathrm{cm}^{-2},
\label{eq:Pc}
\end{equation}
where $c$ is the vacuum speed of light. It follows that the force
between the plates is attractive and monotonically increasing with
decreasing $d$, a result that is often heuristically explained from
the fact that the density of photonic modes is larger outside than
inside the plates.

Because Casimir and classical forces are both related to the mode
frequencies of a system, in principle one can employ classical ideas
from nanophotonics to gain intuition. However, the similarities
between Eq.~\ref{eq:Ucas} and Eq.~\ref{Uopt} in
Section~\ref{sec:opt-theory} belie a number of fundamental differences
that lead to dramatically different physics and design
principles. First, we should note that the use of photon number in the
classical energy expression of Eq.~\ref{Uopt} is merely a heuristic
and convenient way of relating the energy stored in a given mode to
its corresponding frequency, and of course bears no relation to the
physical energy stored in the system, which arises from externally
incident light. (In fact, as mentioned above, one can also derive the
classical force expression without introducing any quantum-mechanical
quantities~\cite{PovinelliJo05}.)  In contrast, the contribution
$\hbar\omega$ of virtual photons in Eq.~\ref{eq:Ucas} has a direct
physical interpretation as arising from zero-point charge (and
electromagnetic) fluctuations inside and outside the bodies, as
opposed to any external sources.  Hence the quantity $\hbar$ appears
in the final expression for the pressure (Eq.~\ref{eq:Pc}). Second,
since the Casimir energy is a result of broad-bandwidth
electromagnetic fluctuations, from electrostatic all the way to UV
wavelengths where materials become transparent, Eq.~\ref{eq:Ucas}
contains a sum over all frequencies. As a consequence, whereas designs
based on classical forces often focus on resonant effects, looking at
the energy or force contributions within a narrow frequency bandwidth
can be very deceptive in the Casimir case.

To illustrate this point, Fig.~\ref{fig:casimag} shows the Casimir
force integrand as a function of $\omega$ for two semi-infinite gold
plates. Note that the expression in Eq.~\ref{eq:Pc} is obtained only
upong taking the idealized limit of a perfect metal ($\varepsilon \to
-\infty$).~\footnote{Technically $U \to \infty$ in closed PEC
  structures, a divergence that this is circumvented in energy
  calculations by exploiting renormalization procedures involving
  high-frequency cutoffs~\cite{milonni,Dalvit11:review}. Such cutoffs
  are motivated by the fact that these divergences are not physical,
  since real materials such as the gold plates of
  Fig.~\ref{fig:casimag} become transparent at large $\omega$.}
Regardless of material considerations one finds that generally, while
certain frequencies give attractive contributions, others give
repulsion, and the overall sign is determined by a complicated
cancellation between these effects.  The sum-over-modes calculation
for dielectric plates dates back to the original work by Van
Kampen~\cite{vanKampen68}, who computed the force in the near field $d
\ll \lambda_p$, where $\lambda_p$ is a characteristic wavelength which
for metals corresponds to the plasma wavelength (typically in the
ultraviolet range). In this quasistatic regime, the force law scales
as $1/d^3$ whereas at large separations $d \gg \lambda_p$, the force
again scales as $1/d^4$. In between, the $d$-dependence of the force
is determined by the frequency-dependent permittivity $\varepsilon$ of
the materials involved. This transition reveals the increasing
contribution of evanescent waves in the near field, as recognized by
Van Kampen. More recently, there has been renewed interest in
analyzing the modal contributions at different frequencies and
separations, which just as in the classical case, can be classified as
arising from either evanescent (gradient) or propagating (radiative)
modes~\cite{Genet03:scat,Enk95,Intravaia05}. Such a decomposition was
explored in Ref.~\cite{Intravaia07}, which found that evanescent modes
not only dominate the interaction at small $d$, but also contribute a
significant repulsive component at larger separations stemming from
the contribution of antibonding modes, with bonding modes contributing
attraction in analogy with classical forces. Ultimately, however, the
sign and scaling of the force is determined by a competition between
the radiative, bonding, and antibonding evanescent modes, which lead
to attraction at all separations. Such an interplay is illustrated in
Fig.~\ref{fig:casimag} for the example of two gold plates. It is worth
noting that, because incident light in classical systems is typically
introduced over a narrow bandwidth, the scaling of gradient forces
with separation is often exponential at large separations $d \gg
\lambda$, where $\lambda$ is the decay length of the corresponding
excited mode. In contrast, Casimir forces arise from contributions
spanning many such modes, each contributing different exponential
decays, the cumulative effect of which leads to the usual polynomial
scaling. Hence, even if the force contributions seem to be
dramatically altered within a narrow bandwidth, as is often the case
in classical systems, this alteration is usually canceled by
contributions at other frequencies and therefore has little influence
on the force.

In addition to posing conceptual challenges, the broad and oscillatory
character of the Casimir spectrum also imposes severe limitations for
theoretical and numerical calculations, as reviewed in
Refs.~\cite{Johnson11:review,ReidRo12:review}. The key to overcoming
these problems is a well-known technique from complex analysis
involving extensions of the real-frequency integrand into the complex
plane, on which practically all modern calculations are
based. Specifically, because the Casimir force integrand is
unobservable and related to causal scattering problems (fields due to
dipole antennas), it is an analytic function in the upper-half complex
frequency plane ($\mathrm{Im}\, \omega > 0$), and hence the integral
along real frequency is mathematically equivalent to the integral
along the imaginary-frequency axis $\omega = i
\xi$~\cite{Lifshitz80}. Intuitively, the presence of oscillations and
resonances in the Casimir integrand is linked to the fact that dipole
sources oscillating at real frequencies $\omega$ produce radially
propagating spherical waves $\sim \frac{\exp(i\omega r/c)}{r}$, where
$r$ is the distance away from the dipole, which in turn lead to the
sensitive interference and cancellation effects employed in classical
designs. Along imaginary frequencies $i\xi$, however, dipole sources
lead to exponentially decaying fields $\sim \frac{\exp(-\xi r/c)}{r}$,
which in turn yields integrands that are smooth, non-oscillatory and
exponentially decaying (decay lenghts $\propto \frac{1}{d}$) and which
are highly susceptible to efficient numerical integration,
e.g. numerical quadrature. The absence of resonances along imaginary
frequencies extends to material responses as well, since the
permittivity of passive materials $\varepsilon(i\xi)$ evaluated at
imaginary frequencies decays monotonically with
$\xi$~\cite{Lifshitz80}.  To illustrate this exponential behavior,
Fig.~\ref{fig:casimag}(b) shows the Casimir integrand between gold
plates along the imaginary frequency contour. While this
imaginary-frequency perspective is crucial for numerics, it also
suggests that when thinking of modifications to Casimir forces (in
contrast to classical design principles that focus mainly on resonant
phenomena) it is particularly useful to employ intuition from
quasistatics ($\omega \to 0$), where the lack of resonances and
interference effects captures the decaying, non-oscillatory physics at
imaginary frequencies~\footnote{Mathematically, the exponential decay
  of the Casimir-force contributions at imaginary frequencies
  $\omega=i\xi$ tends to make the total force qualitatively similar to
  the $\xi=0$ contribution, which dominate in the near-field or van
  der Waals regime.}.


\subsection{Designer Casimir forces}
\label{sec:cas-designs}

Unlike classical forces, the study of Casimir forces was until
recently hampered by a lack of theoretical tools.  Accurate
experimental measurements were also until recently extremely
challenging due to the small magnitude of these interactions at large
separations $\gtrsim \mu$m. The emergence of new theoretical tools and
state-of-the-art experimental setups is finally enabling synergistic
explorations of this phenomenon, and one emerging topic of interest is
the design of structures exhibiting force behaviors that differ from
the standard attractive $1/d^4$ powerlaw of planar PEC bodies, and
which aid in reducing or altogether switching the sign of the
force. As discussed below, a few of these new theoretical predictions
have been realized in experiments, though the range of potential
effects remains largely unstudied.

\subsubsection{Theory}

The sum-over-modes approach described above provides a conceptually
useful starting point for understanding Casimir forces and has been
subject to multiple modifications to include realistic structures
including lossy materials and finite
objects~\cite{vanKampen68,Gerlach71,milonni,Genet03:scat,Enk95,Graham09,Mochan06,Davids10}. However,
it turns out that in practice it is seldom employed for calculations,
which more commonly rely on modern perspectives and theoretical
techniques that lie much closer to the original picture of fluctuating
dipoles introduced by van der Waals and
others~\cite{lifshitz1,Dzyaloshinskii61}. In particular, this
intuitive picture provides a direct and intuitive connection to
classical scattering problems by way of the fluctuation-dissipation
theorem of statistical physics~\cite{lifshitz1,Eckhardt84,Lifshitz80},
relating the dissipative properties of bodies to the resulting dipole
fluctuations. Although such a relationship was first derived in the
context of microscopic dipole--current fluctuations within the volumes
of the bodies, more recent approaches reduce the problem to a smaller
number of scattering problems involving sources and/or incident
electromagnetic waves only on surfaces around
bodies~\cite{Johnson11:review}. In fact, the first generalization of
Casimir's original derivation was carried out by Lifshitz,
Dzyaloshinski{\u{\i}}, and Pitaevski{\u{\i}} in the
1960s~\cite{Dzyaloshinskii61} where, rather than summing mode
contributions, they considered instead calculations based on
thermodynamic Maxwell stress tensors (ensemble average of
Eq.~\ref{eq:Tij}) and fluctuation--dissipation theorems of fields at
equilibrium, which relate expectation values of fluctuating fields
$\langle E_i E_j \rangle$ to the imaginary part of the electromagnetic
Green's function (electric field due to a dipole
source)~\cite{Lifshitz80}.~\footnote{Note that in contrast to
  classical calculations, there are no ambiguities when evaluating
  energy or stress tensors due to fluctuations in dissipative media
  since in the Casimir case one is dealing with thermodynamic
  quantities~\cite{Pitaevski06,Philbin11,Reid12:FSC}.}  The
stress-tensor approach was originally employed as a semi-analytical
method to obtain predictions in simple geometries, such as planar
dielectric bodies and eventually arbitrary multilayer
films~\cite{Jaekel91,Zhou95,Klimchitskaya00,Tomas02}.

For the next few decades, virtually all theoretical work focused on
studying ways of altering the magnitude and sign of the force by
employing either semi-analytical techniques or heuristic
approximations involving small perturbations around known solutions,
such as parallel plates~\cite{Derjaguin34,bordag01} or dilute
gases~\cite{Milton08,Milton08:scat,Golestanian09}. As described in
more detail in Section~\ref{sec:cas-designs}, approaches based on
materials in planar structures relied mainly on narrow resonance
effects whose contributions wash out when the full spectrum is
considered~\cite{Rosa09,McCauleyZh10}. Modification arising from
boundary effects were also studied, albeit using ad-hoc approximations
valid only in the limit of small curvatures. These include pairwise
additive approximations such as the well-known Derjaguin or
proximity-force approximation
(PFA)~\cite{Derjaguin56,Klimchitskaya00,Bordag06}, which to lowest
order yield the force between nearly planar bodies by summing the
force per unit area of an equivalent parallel-plate geometry over all
adjacent surfaces, providing an intuitive way of conceptualizing
changes to the force. For instance, within PFA the zero-temperature
Casimir force between a PEC sphere of radius $R$ and a PEC plate
separated by distances $d \ll R$ is given by $F_\mathrm{C} =
-\frac{\hbar c \pi^3 R}{360 d^3}$~\cite{Derjaguin34}. Similar
closed-form expressions can be worked out for a handful of
high-symmetry structures, including infinite cylinders above
plates~\cite{emig06}, in which case the force per unit length
$p_\mathrm{C} = -\sqrt{\frac{R}{2d}} \frac{\hbar c \pi^3}{384 d^3}$,
with the main result being a change in the power-law of the
interaction. In situations involving dilute or weakly polarizable
media, approximations based on summation of van der Waals or
Casimir--Polder interactions can be
employed~\cite{Milton08:scat,Golestanian09,Bitbol13}. Other
perturbative approximations include re-normalized
Casimir-Polder~\cite{Tajmar04, Sedmik06} or semi-classical
interactions~\cite{Schaden98}, multiple scattering
expansions~\cite{Balian78,Emig03,Lambrecht06,Maghrebi10:expansion},
classical ray optics approximations~\cite{Jaffe04,Jaffe05:approach},
and higher-order PFA corrections~\cite{Bordag06}.

When applied to non-planar structures with separations $d \sim$
feature sizes and in non-dilute media, these approximations fail to
capture fundamental non-additive
effects~\cite{Rodriguez11:review,Bitbol13} that can only be handled by
more sophisticated techniques. Mirroring the development of classical
electromagnetic calculations, early approaches mainly considered
semi-analytical methods that expoit symmetries in specific geometries
to expand the field unknowns in convenient spectral bases, such as
Fourier modes or spherical
multipoles~\cite{Balian78,Mazzitelli06,Dalvit06,Lambrecht06,Emig07,Milton08,Kenneth08,Milton08,Lambrecht09,Rahi09:PRD,Davids10}.
It took almost four decades for sophisticated numerical techniques
that can capture the full spectrum of non-additive effects in
realistic, microstructures geometries to be
developed~\cite{Johnson11:review,Rodriguez11:review}. Such
general-purpose methods come in a bewildering variety of flavors,
including formulations based on the original stress tensor approach,
fundamentally tied to the statistics of
fluctuations~\cite{Lifshitz80,Rodriguez07:PRA}, as well as more modern
approaches in which the force is derived or expressed via path
integrals~\cite{gies03,emig03_2,emig04_2,Gies06:worldline,Rahi09:PRD,Reid11}
or trace formulas involving scattering
matrices~\cite{Lambrecht06,Emig07,Kenneth08,Rahi09:PRD,MiltonPa10}. Similar
to classical calculations, numerical implementations vary and include
spectral-basis or eigenmode expansions, e.g. Fourier series and
spherical harmonics)~\cite{Emig07,Davids10}, or localized mesh
discretizations, e.g. finite
differences~\cite{Rodriguez07:PRA,Pasquali08,Pasquali09,RodriguezMc09:PRA,McCauleyRo10:PRA}
or boundary
elements~\cite{ReidRo09,Xiong09,Xiong10,Reid11,Reid12:FSC,Atkins13}. Despite
their deceivingly different origins and algebraic structures, the
thread connecting all of these formulation is the underlying
relationship between electromagnetic fluctuations and classical
scattering problems. Because many such scattering problems must be
solved to compute even a single force, calculations tend to be
significantly more computationally demanding than classical force
simulations. However, as discussed further below, despite these
complications the development of exact methods has ushered a
renaissance of Casimir-force modeling in highly non-planar
structures. For more in-depth discussions of theoretical methods
emphasizing the connection between classical calculations and Casimir
forces, the reader is referred to
Refs.~\cite{Johnson11:review,ReidRo12:review}.

\subsubsection{Experiments}
\label{sec:cas-exps}

\begin{figure*}[th!]
  \centering \includegraphics[scale=0.575]{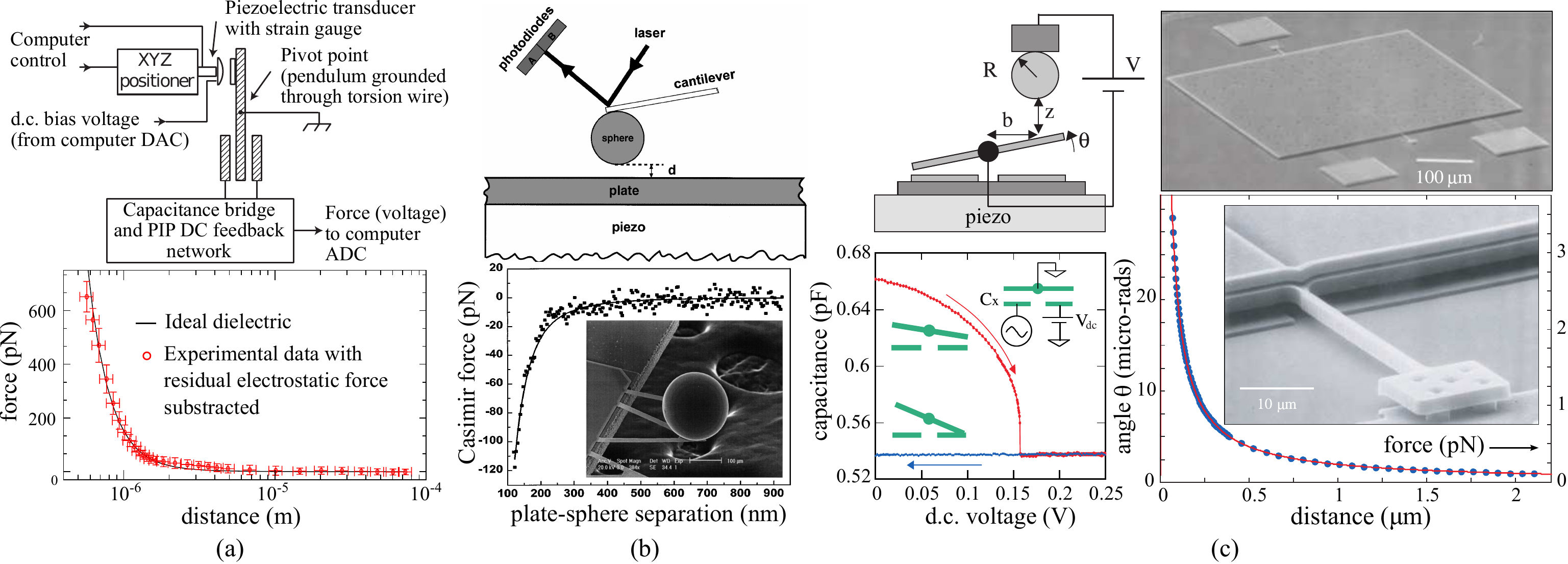}
   \caption{Examples of various experimental setups for detecting
     Casimir forces, including: (a) Torsional pendulums where the
     static change due to the Casimir force between a sphere and a
     plate is measured as a change in the capacitance between two
     compensator electrodes.  The setup and corresponding force
     measurements are taken from
     Refs.~\cite{Lamoreaux97,Kim2009,Sushkov10}, and involve forces
     between a 1" gold-coated quartz plate and a gold-coated spherical
     lens with a radius of curvature of 11.3cm.  The experimental data
     show the Casimir force with the residual electrostatic force
     subtracted from the total force measurements. (b) AFM setups
     where both the static and dynamic modes have been utilized for
     measuring the deflection and mechanical frequency-shift of a
     cantilever~\cite{Mohideen98,Mohideen2000,Chang2011,Banishev14,Torricelli2010,Torricelli2011,Laurent2012,Chang2012,Banishev2012,Banishev13:ni,Banishev13:graphene}.
     The experimental data is taken from Ref.~\cite{Mohideen98}, and
     shows force measurements obtained by calibrating the cantilever
     deflection signals.  The inset illustrates the corresponding
     experimental apparatus, involving a metallized polystyrene sphere
     with a diameter of 0.2mm attached to an AFM tip. (c) Torsional
     MEMS oscillators driven capacitively by electrodes, where the
     frequency shifts due to the Casimir force can be measured
     electrically or optically. The setup and measurements are taken
     from Ref.~\cite{hochan1}, involving a doped polysilicon 500$\mu$m
     $\times$ 500$\mu$m plate anchored to a substrate and acting as a
     torsional oscillator. The angular displacement $\theta$ of the
     oscillator is obtained by measuring the capacitance change
     between the oscillator and the electrodes below, which involves
     biasing one of the electrodes and application of a small AC
     probing signal.  The angular displacement at various sphere-plate
     separations is then normalized to obtain the force data.}
   \label{fig:casexps}
\end{figure*}



The first experimental validations of the Casimir effect date back to
the 1960s, although significant sources of errors made it difficult to
unambiguously confirm the existence of the force~\cite{Sparnaay58}. In
early experiments, complications arose from the need to maintain a
high degree of parallelism between plates and from the small force
magnitudes present at large separations. Decades later, advances in
instrumentation began making it possible to place nearby objects in
close proximity with nanometer precision and slow mechanical drifts,
facilitating measurements with significantly higher precisions and
spatial resolutions. It wasn't until the experiments of van~Blokland
et. al. that the force between metallic surfaces was unambiguously
measured~\cite{Overbeek78}, followed almost two decades later with the
first high-precision measurements of the Casimir force~\cite{lamo},
ushering a new era of Casimir metrology. In order to circumvent
alignment issues related to parallelism between planar surfaces, most
Casimir-force measurements have since been conducted in a related
geometry involving a large (tens of microns or more) sphere in close
proximity to a planar body. Among the few exceptions was the
experiment by Bressi et. al. in 2002, which unmambiguously verified
the Casimir force in the original parallel-plate
geometry~\cite{bressi}. Readers are referred to
Refs.~\cite{bordag01,Dalvit11:review,Lamoreaux05} for a historical
overview of Casimir experiments.

In most experiments, the Casimir effect manifests itself as a force
that shifts the equilibrium position, or alternatively which modifies
the spring constant of a compliant test body when in close proximity
to a surface, similar to other distance-dependent forces like the
aforementioned optical and electrostatic forces.  To date, most
observations rely on measuring changes in the equilibrium position of
the object in question (a static effect) or force gradients (a dynamic
effect) over some range of separations.  Due to difficulties involving
large, environmental DC noise, static measurements are less frequent
than dynamic measurements of the force gradient. In what follows, we
briefly describe three common sets of experimental setups, illustrated
in Fig.~\ref{fig:casexps}, based on either torsional pendulums,
dynamic-mode atomic force microscopy (AFM) or torsional MEMS
oscillators.

The torsional pendulum setup developed by the Lamoreaux group, shown
in Fig. \ref{fig:casexps}(a), was employed in the first high-precision
(5\% accuracy) demonstrations of the Casimir effect and more recently
in one of the first demonstrations of the thermal Casimir force
\cite{Lamoreaux97,Kim2009,Sushkov10}. In such an experiment, the
static change due to the force exerted on two bodies (a flat,
metal-coated plate and a large, metal-coated spherical lens) is
directly measured in vacuum and at room temperature. While one end of
the sphere is mounted on the far end of a pendulum, the other is
sandwiched between two nearby compensator electrodes that enable the
sphere to move. Specifically, as these two bodies are brought closer
together using a closed-loop piezoelectric transducer, the attractive
Casimir force exerts a force on the sphere and a corresponding torque
on the pedulum, causing an imbalance in the capacitance of the two
compensator electrodes.  A corresponding voltage is sent to one of the
electrodes to counteract the effect of the torque and therefore
restore the capacitance balance with the two electrodes, where the
error signal detected and applied is directly proportional to the
static magnitude of the Casimir force.  With this particular
apparatus, accurate measurements can often be made at separations
spanning the range 0.7-7$\mu$m.

A related scheme involves static and dynamic AFM measurements of the
separation of a sphere attached to a cantilever from a surface. In the
static mode, the displacement is obtained by measuring the deflection
of the cantilever tip with a position-sensitive
detector~\cite{Mohideen98, Mohideen2000,Chang2011,Banishev14}.  For
higher detection sensitivities, both the amplitude- and
frequency-modulation modes of AFMs can be used to detect the frequency
shift of the cantilever. The amplitude-modulation mode works by
driving the cantilever near a resonance frequency, which leads to
appreciable amplitude changes in the detected signal even under small
frequency shifts. Experiments adopting this measurement mode include
those of Refs.~\cite{Man09,Man10:PRA,Jourdan09}.  The
frequency-modulation mode of an AFM, on the other hand, offers higher
sensitivity than the amplitude-modulation mode and is also more immune
to amplitude noise. In this case, the cantilever is driven on
resonance and the resonant frequency is monitored with a fiber
interferometer and tracked with a phase-lock loop, as illustrated in
Fig.~\ref{fig:casexps}(b). The frequency shift due to short-range
attractive forces can then be obtained by frequency demodulation: it
is brought close to the other test body by a closed-loop piezoactuator
where the test body's position is monitored by another fiber
interferometer (with sub-nm resolution). Reported achieved separations
in this setup range from 200nm to
2$\mu$m~\cite{Torricelli2010,Torricelli2011,Laurent2012,Chang2012,Banishev2012,Banishev13:ni,Banishev13:graphene}.

Finally, the torsional MEMS oscillators employed by the Chan and Decca
groups have proven useful for performing measurements at the
microscopic
scale~\cite{hochan1,hochan2,decca1,Decca2003,Chan08,Intravaia13}.  As
illustrated in Fig.~\ref{fig:casexps}, the torsional oscillator is
often a large metal or heavily doped semiconductor (e.g. polysilicon)
plate suspended by thin arms and engineered to carefully tune its
mechanical resonance frequency. Attached to the plate lies one of the
test bodies, often a large-diameter sphere, while underneath lie two
electrodes, driven by resonant and modulation signals in order to
detect the capacitance change between the two, and hence the force
gradient. In contrast to other setups, here calibration of the
absolute separation between the test bodies can be performed by
electrostatic or optical interferometric means~\cite{Zou13}.

Beyond parallelism, challenges to high precision experiments include
electrostatic forces that arise due to either static potentials
between different metals connecting the two surfaces, or patch
potentials~\cite{Lamoreaux10:review,Wei10,Man10,GarciaSanchez12}. Electrostatic
effects can be the dominant interaction between surfaces at large
separations, and hence minmizing and accounting for their contribution
is crucial for obtaining accurate measurements at $\mu$m
lengthscales. In measurements where the Casimir effect reduces the
mechanical spring constant of the test bodies, the typical protocol
for calibrating these is by measuring the corresponding electrostatic
force as a function of the applied voltage. On the other hand, patch
potentials are highly sample dependent and cannot easily be altered in
situ, and thus their impact is often fitted via empirical
models~\cite{Speake2003,Bordag09:book,Kim10,Behunin2012}. These
effects can become even more pronounced in polycrystalline metal
films~\cite{Behunin2012b}. The Tang group recently developed an
\emph{in situ} scanning Kelvin probe, involving a $1\mathrm{mm}\times
1\mathrm{mm}$ metallized silicon nitride membrane above an 8mm gold
sphere, which enables measurements of the spatial distribution of such
residual potentials, allowing in-situ detection of force
gradients~\cite{GarciaSanchez12,Garcia2013} and providing a valuable
technique to ascertain the role of such patch potentials on
integrated, Casimir force measurements.

Mature nanofabrication and metrology techniques have begun to greatly
influence force measurements, enabling careful control and calibration
of material and surface properties. First, while film deposition and
microfabrication of MEMS devices are straightforward, evaluating the
influence of surface roughness to the Casimir effect at short
distances demands care in refining the fabrication procedures against
typically tolerable imperfections, including surface roughness and
existence of native oxides
\cite{ederth,genet03,Zwol08:rough,Broer12,Chan2010}.  The metals used
in Casimir experiments are typically prepared by evaporation or
sputtering in ultrahigh vacuum environment and at low deposition
rates.  Recent AFM measurements have been employed to survey the
surface morphology, root-mean-squared (rms) roughness and peak-to-peak
height variations of various metallic samples, leading to better
descriptions of roughness at short lengthscales. For instance, it was
found that while typical rms roughness is on the order 2nm,
peak-to-valley height variations could reach tens of
nanometers. Theoretical models that incorporate such detailed
descriptions of roughness are discussed in
Section~\ref{sec:cas-progress}. Accurate comparison between theory and
experiments also requires accurate knowledge of $\mathrm{Im}\,
\varepsilon$ over a wide spectral
region~\cite{Torricelli2010,Torricelli2012}, and variations and
uncertainty among samples used in different experiments can also prove
problematic. Recent efforts characterizing the dielectric spectral
properties of materials under test by ellipsometry have begun to shed
light on a number of these
issues~\cite{Pirozhenko06,Svetovoy2008,Zwol09,Zwol10,Olmon2012,Sedighi14}. Finally,
demonstrations of switchable Casimir forces with phase-change material
such as $\mathrm{Ag_5In_5Sb_{60}Te_{30}}$ (AIST), which can turn from
amorphous to single-crystalline state by annealing
\cite{Torricelli2010:phase,Torricelli2012}, are paving the way for the
flourishing development of versatile operations where the Casimir
force can be controlled \emph{in situ} with fast response times, and
this is bound to play an impact on the accuracy of future experiments.

The recent push to measure Casimir forces in unconventional, complex
geometries and the development of compatible computational tools is
also beginning to encourage application of more novel fabrication
techniques. Specifically, there has been growing interest in
developing integrated systems for probing and even utilizing the
Casimir force~\cite{Zou13,RodriguezWo11,Pernice10}.  These on-chip
schemes remove the need for external instrumentation needed to bring
test bodies together which are often prone to mechanical drifts.  In
fact, microfabrication could provide a means of eliminating the
experimental challenge of attaining high degree of parallelism for
measurements in the plate-plate
geometry~\cite{Nawazuddin2010,RodriguezWo11}. A particular design
proposed and fabricated by the MEMS community is a silicon plate
tethered by three surrounding electrostatic
actuators~\cite{Yamarthy2009}, overcoming the need for direct
actuation of test bodies which otherwise introduces additional spring
effects into the measurement.  A more recent demonstration by the Chan
group employed a comb drive actuator based on a highly doped
silicon-on-insulator platform that was used to indirectly introduce a
movable doubly clamped beam as a force-sensing element~\cite{Zou13},
illustrated in Fig.~\ref{fig:casgeom}(c).


\begin{figure*}[tb!]
  \centering
  \includegraphics[scale=0.27]{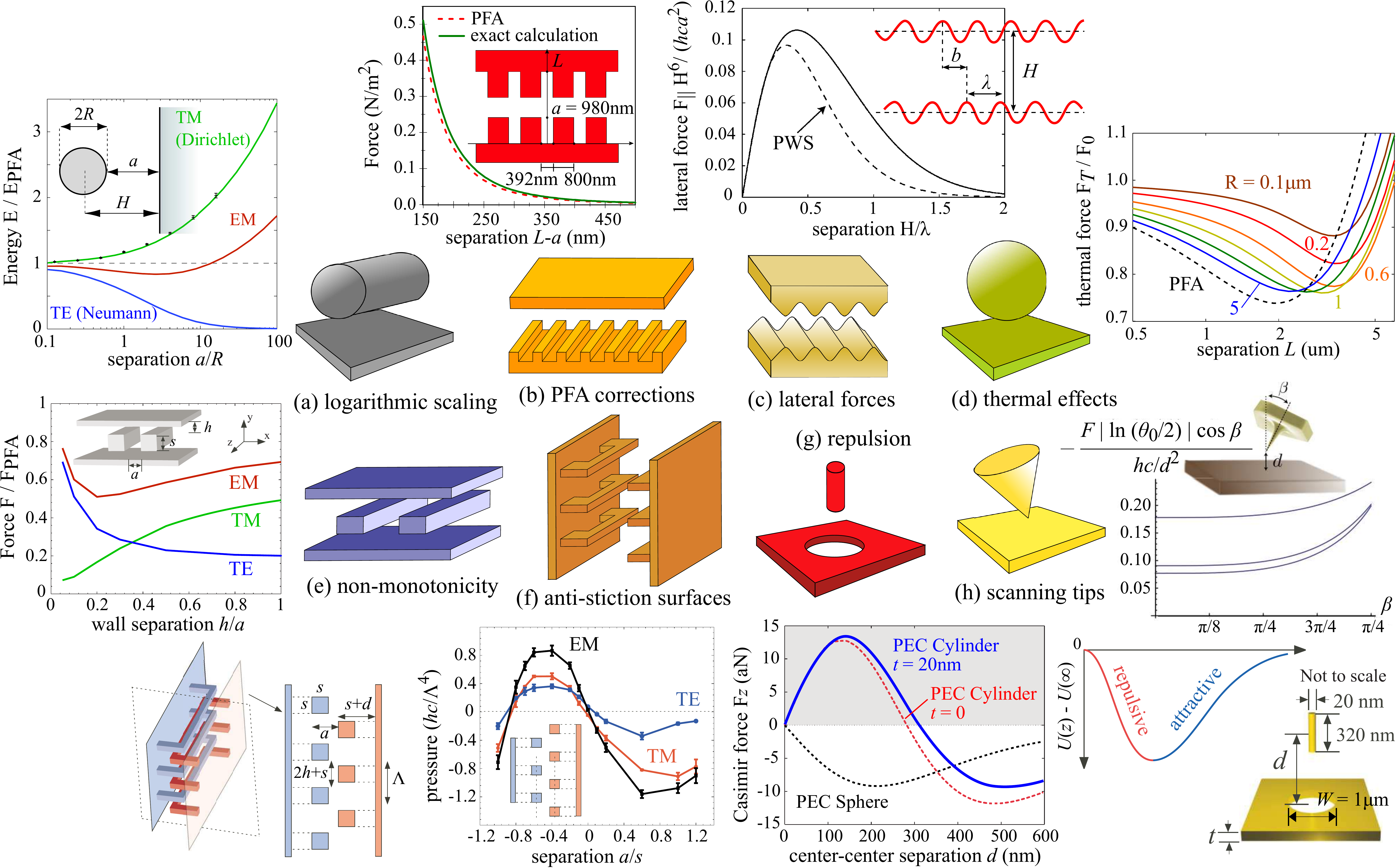} \caption{Selected
    theoretical structures along with predictions of unusual Casimir
    effects. (a) Energy per unit length of a PEC cylinder of radius
    $R$ above a PEC plate as a function of separation $H$, normalized
    by the corresponding PFA energy and decomposed into both
    transverse-electric (TE, blue line) and transverse-magnetic (TM,
    red line) solutions~\cite{emig06}. It is found that the total
    force (red line) in this configuration has one of the weakest
    decays $\sim H^{-3} \log(R/H)$ as $H/R \to \infty$. (b) Pressure
    between two silicon rectangular gratings, computed by exact
    (green) and PFA (red) methods. (c) Normalized lateral pressure
    between two sinusoidally corrugated PEC surfaces of period
    $\lambda$ as a function of their surface--surface separation $H$,
    computed by both exact methods (solid) and pairwise summation
    (PWS) approximations (dashed)~\cite{emig03_1}. The lateral force
    is shown to be maximized at an optimal periodicity $\lambda
    \approx 2.5H$, and the PWS is shown to become progressively worse
    with increasing $H/\lambda \gtrsim 0.3$. (d) Thermal Casimir force
    $F_T$ between a gold sphere and plate at room temperature, divided
    by the force at zero temperature $F_0$, as a function of their
    separation $L$. Solid lines from top to bottom correspond to
    increasing values of sphere radius $R$. The dashed curve
    represents the same force ratio, but as computed by PFA. (e) Force
    per unit length between two PEC waveguides adjacent to two PEC
    plates a distance $h$ apart, normalized by the corresponding PFA
    force and decomposed into both TE (blue line) and TM (green line)
    solutions~\cite{Rodriguez07:PRL}. It is shown that the competition
    and opposite behavior of TM and TE forces with respect to changes
    in $h$ causes the total force (red line) to have a non-monotonic
    dependence on $h$. (f) Pressure between two interleaved PEC
    zipper-like structures (with cross-section shown on the inset)
    involving brackets attached to plates, decomposed into both TE
    (blue line) and TM (red line) solutions, as a function of the
    center--center separation $a$ between the
    brackets~\cite{RodriguezJo08:PRA}. The competition between lateral
    and normal forces leads to a change in the sign of the force,
    becoming repulsive at an intermediate separation due to lateral
    attraction between the brackets. (g) Casimir force between a small
    cylindrical particle and a plate with a hole of size $W$ as a
    function of their center--center separation $d$, for both gold
    (blue) and perfect metal (dashed red) material
    configurations~\cite{LevinMc10}. The force becomes repulsive at
    small separations $d \lesssim W/3$, which is explained from the
    fact that (by symmetry) the energy of a small $z$-oriented dipole
    located at the center of an infinitesimally thin plate with a hole
    is equal to the energy of an equivalent, isolated dipole. (h)
    Thermal Casimir force between a tilted PEC cone suspended above a
    PEC plate by a distance of $1\mu$m, as a function of the tilt
    angle $\beta$, for multiple temperatures $T=300,80$, and $0K$ (top
    to bottom).}  \label{fig:cas-collage}
\end{figure*}

\subsubsection{Recent developments}
\label{sec:cas-progress}

\begin{figure*}[tb!]
  \centering
   \includegraphics[scale=0.42]{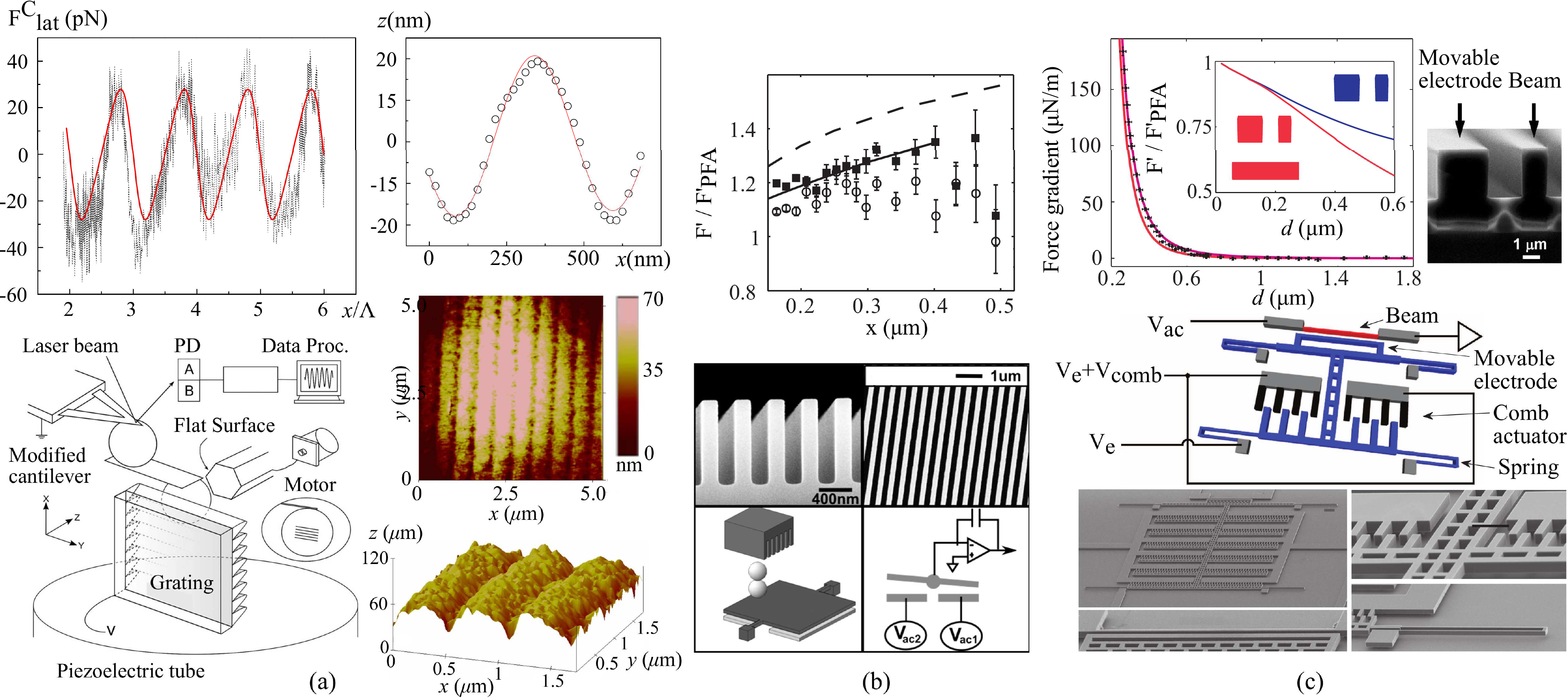}
   \caption{Recent experimental demonstrations of lateral forces,
     deviations from PFA, and finite-size effects in microstructured
     geometries. (a) Schematic of experimental setup used to measure
     lateral Casimir forces (i), involving an AFM system comprising
     two test bodies, a polystyrene sphere of diameter 200$\mu$m and a
     plate, whose surfaces are covered with longitudinal corrugations
     of average period $\Lambda\approx 574.7$nm~\cite{Chiu10}. Also
     shown are AFM scans (ii,iii) of the corrugated surface of the
     sphere along with a cross-section (iv) of the grating surface,
     where the solid line shows a sine function obtained by
     fitting. The plot (v) shows the measured lateral force (black
     dots) as a function of lateral displacement $x$ for a single
     separation of $124.7$nm, whose asymmetric pattern shows good
     quantitative agreement with exact theoretical predictions (red
     line). (b) Lateral (i) and top (ii) cross-sections of a
     rectangular trench structure with periodicity $\Lambda=400$nm and
     depth $t=0.98\mu$m, suspended above two stacked silica spheres of
     diameter $50\mu$m mounted on a movable silicon plate
     (iii,iv). Plotted is the ratio $F'/F'_{PFA}$ of the measured
     Casimir force gradient between the top sphere and the grating,
     normalized by the corresponding PFA force gradient, as a function
     of their surface--surface separation $d$, for samples with
     $2\Lambda/t=1.87$ (circles) and $2\Lambda/t=0.82$ (squares). Also
     plotted are theoretical predictions based on exact
     methods~\cite{Lambrecht09}, which demonstrate strong deviations
     from PFA. (c) Schematic of an integrated experimental setup (i)
     along with corresponding scanning-electronic micrographs (ii,iii)
     of a structure involving a silicon beam (red) and a movable
     silicon electrode and comb actuator supported by four springs
     (blue), with electrical connections~\cite{Zou13}. The plot (iv)
     shows the measured Casimir force gradient between the beam and
     electrode (black dots) as a function of separation $d$, along
     with exact theoretical calculations in the absence (red line) and
     presence (purple line) of electrostatic patch potentials. The
     inset shows the calculated force normalized by the expected PFA
     force in the presence (red) and absence (blue) of the substrate,
     showing that multi-body effects become appreciable at separations
     $d\gtrsim 2\mu$m.}
   \label{fig:casgeom}
\end{figure*}

\begin{figure}[tb!]
  \centering
   \includegraphics[scale=0.6]{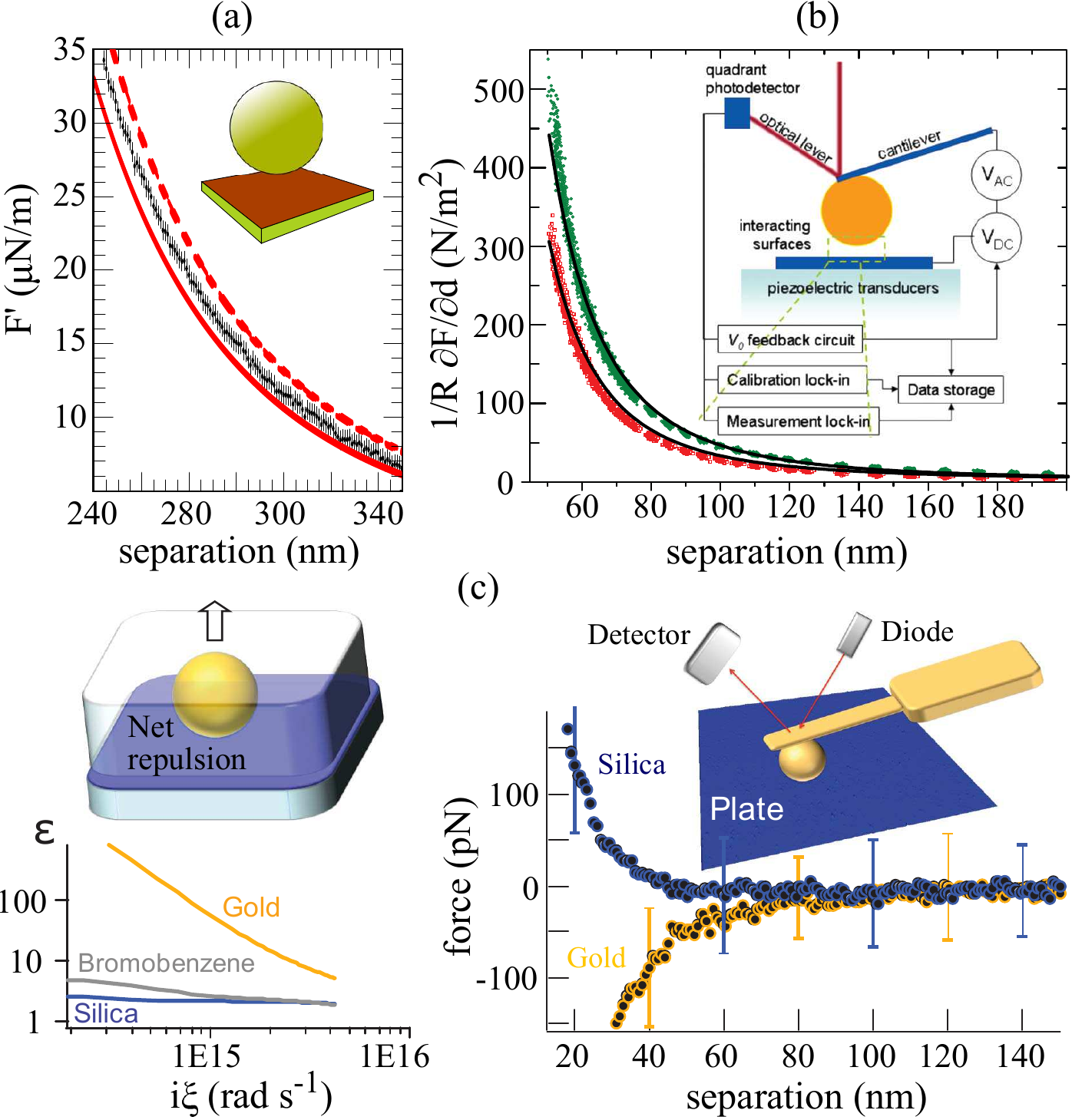}
   \caption{Recent experimental demonstrations of modifications to
     Casimir forces due to material effects (all involving AFM
     setups). (a) Measured force gradient between a silica sphere of
     radius $54\mu$m, coated with 280nm gold film and suspended above
     a graphene-coated silicon-on-insulator substrate (black crosses),
     along with corresponding predictions from the Lifshitz theory in
     the absence (solid line) and presence (dashed line) of the
     graphene sheet~\cite{Banishev13:graphene}. The graphene sheet is
     shown to increase the Casimir force by $\approx 10\%$.  (b)
     Measured gradient force between a gold-coated polystyrene sphere
     of radius $R=100\mu$m and gold thickness $\sim 100$nm and either
     a gold-coated (green square) or ITO-coated (red squares) silica
     plate, as a function of their surface--surface separation
     $d$~\cite{Man09}. The black lines indicate calculated Casimir
     forces with the addition of residual electrostatic forces. The
     use of oxides is shown to halve the Casimir force by roughly
     50\%. (c) Measured attractive (orange) and repulsive (blue)
     Casimir forces between a gold sphere and either a silica or gold
     plate, respectively, immersed in
     bromobenzene~\cite{Munday09}. Also plotted are the dielectric
     permittivities $\varepsilon$ of gold (orange), bromobenzene
     (gray), and silica (blue) evaluated at imaginary frequencies
     $i\xi$, whose dielectric arrangement over a wide range of $\xi$
     leads to repulsion~\cite{Dzyaloshinskii61}.}
   \label{fig:casmat}
\end{figure}

In the last two decades, there have been a plethora of designs and
experiments aimed at modifying forces in microstructured devices,
spurred for example by the desire to reduce stiction in MEMS.  A
commonality of current efforts is the search for structures that
exhibit strong deviations from the additivity inherent in pairwise
approximations such as PFA. Other efforts focus on discovery of
structures/materials exhibiting strong material/temperature
effects. Recent predictions include strong interplay between geometry
and temperature effects in sphere--plate geometries, lateral forces
between corrugated plates, non-monotonic forces between waveguides,
reduced forces between rough surfaces, and even repulsive forces
arising from geometry alone. A selection of representative structures
and predictions of unusual behaviors is shown in
Fig.~\ref{fig:cas-collage}. Figures~\ref{fig:casgeom}
and~\ref{fig:casmat} describe recent experiments exploring force
modifications based on either geometrical or materials approaches.

The first investigations of non-additive or unusual geometric effects
considered idealized situations that proved to be theoretically
convenient but experimentally unrealistic. These include predictions
of repulsive or expansive self-forces in hollow PEC and
exotic-dielectric shells~\cite{Boyer68,Milton78,Brevik88,bordag01} or
``Casimir pistons'' consisting of two PEC blocks sliding between PEC
walls~\cite{Cavalcanti04,Hertzberg07,Marachevsky08}, which later
proved problematic. For instance, the repulsive effect was shown to
disappear if the sphere is cut in half~\cite{Kenneth02} and also goes
away when realistic materials are considered in which case the Casimir
energy is highly dependent on artificial, high-frequency
cutoffs~\cite{Jaffe03,Dalvit11:review}. Repulsive forces on PEC boxes
were also invalidated when the box expansions were replaced by the
rigid motion of a ``piston'' sliding through a
wall~\cite{Hertzberg07}. A second wave of predictions and experiments
considered more realistic structures, starting with the convenient
sphere--plate configuration.

Despite the commonality of sphere-- and cylinder--plate geometries in
experiments, exact predictions in these systems were obtained only
recently. A state of the art calculation in 2006 was a semi-analytical
calculation by Emig. et. al. in which the force between an infinite
PEC cylinder above a PEC plate was computed via path integral
techniques~\cite{emig06}. The results [Fig.~\ref{fig:cas-collage}(a)]
demonstrated an unexpectedly weak, logarithmic decay $\sim d^{-3}
\log(R/d)$ at large cylinder--plate separations and provided the first
exact prediction of finite temperature contributions (beyond quantum
fluctuations) in an open geometry. Ironically, it took a few years for
the sphere--plate geometry to be explored by exact
methods~\cite{gies06:PFA,Neto08,Emig08:sphpl}, demonstrating
departures from PFA at unexpectedly short separations. Variations of
these structures to include interacting (even eccentric) cylinders,
spheres and ellipsoidal bodies, would eventually be studied as
well~\cite{Dalvit06,Rahi07,Emig07,Neto08,Emig09:ellipsoids}. An
interesting geothermal effect involving nontrivial interplay between
geometry and temperature was recently predicted in the sphere--plate
geometry, wherein thermal fluctuations on the scale of the thermal
wavelength lead to strong departures from PFA and even nonmonotonic
behaviors [Fig.~\ref{fig:cas-collage}(d)] below certain critical
temperatures~\cite{Neto08,Weber10,Weber10:geothermal}. The interaction
between sharp-shaped conductors, such as wedges and cones or the tip
of a scanning tunneling probe, and planar substrates was also recently
explored [Fig.~\ref{fig:cas-collage}(h)], and shown to lead to
significant temperature corrections~\cite{Maghrebi10}. No experiments
have yet probed these geometric effects.

Calculations and experiments demonstrating non-additive effects in
extended geometries include sinusoidal corrugated and grating
structures, starting with the early experiments of Roy
et. al.~\cite{roy99}.  Perturbative expansions in the separation and
corrugation depth of the structures proved useful as they lead to
constraints on the range of validity of PFA in these systems, but soon
had to be extended to capture wavelength-scale
features. Ref.~\cite{Buscher04} considered PEC gratings beyond PFA by
employing path-integral and scattering matrix techniques,
demonstrating $\sim 20$\% deviations from PFA predictions in
structures with periods $\sim$ separations. A recent experiment by
Chan et. al. measured forces in a similar, albeit realistic structure
consisting of a gold sphere above a silicon nanoscale rectangular
grating [Fig.~\ref{fig:casgeom}(b)]~\cite{Chan08}, which was
subsequently followed up by theoretical calculations
[Fig.~\ref{fig:cas-collage}(b)] accounting for changes from the finite
permittivities of the materials~\cite{Lambrecht09}. Comparison of the
experimental measurements with the PEC predictions of
Ref.~\cite{Buscher04} are shown in Fig.~\ref{fig:casgeom}(b) along
with the experimental setup, which consisted of force-gradient
measurements obtained via shifts in the oscillation frequency of the
torsional oscillator on which the sphere was
mounted. Fig.~\ref{fig:cas-collage}(b) shows the theoretical
predictions by Guerout et. al. which include material effects and
confirm deviations from PFA on the order of 10--20\%. Extensions of
these studies to similar and even 2d-periodic
structures~\cite{Davids10,Guerout13} have been explored, demonstrating
strong interplay between geometry, material, and temperature effects;
there are even recent claims of dramatic force reductions (on the
order of 80\%) in gratings with sub-100nm features~\cite{Intravaia13},
although the source of such a reduction has yet to be clearly
identified.

In addition to normal forces, it is also possible to engineer lateral
Casimir forces that act tangentially to the interacting surfaces,
e.g. by breaking translational symmetry. The lateral force between
sinusoidally corrugated surfaces was among the first realistic
structures to be studied using exact
methods~\cite{emig01,emig03_2,Rodrigues06}, leading to predictions of
optimal modulation lengths at which lateral forces are maximized
[Fig.~\ref{fig:cas-collage}(c)]. Applications of lateral forces to
noise-driven ratchets~\cite{Emig07:ratchet} and non-contact
pinions~\cite{Golestanian08} were subsequently explored due to their
potential impact on MEMS. The first experimental demonstration of such
a lateral force was performed in 2002 by Chen et. al. using an AFM
setup comprising a sphere adjacent to a flat plate with nanoscale
sinusoidal corrugations of amplitudes $\sim 5$--$60$nm and periods
$\lambda \approx 1.2\mu$m much larger than the separation between the
bodies. Figure~\ref{fig:casgeom}(a) shows measurements and a schematic
of a similar setup from a more recent experiment~\cite{Chiu10} based
on structures with smaller periods, $\lambda \approx 0.5\mu$m, and
larger corrugation depths, in which larger deviations from PFA and
asymmetric dependencies on the lateral displacement between the plates
were observed. Using a related setup, the same authors measured the
dependence of the force as a function of the angle between the
corrugations~\cite{Banishev14}.  A related structure consisting of
Ni-corrugated surfaces, one of which is hidden under a thin metallic
film, has also attracted recent interest~\cite{Bimonte14} due to
predictions of strong variations in the normal force with the lateral
positions of the surfaces and its corresponding sharp dependence on
the choice of metal. Such a system would allow unambiguous
investigations of the role of dissipation on the thermal Casimir
force, a subject of recent controversy~\cite{milton04,Lamoreaux05}.

Progress in modeling nanostructured surfaces has also led to
significant breakthroughs in understanding the role of surface
roughness (disordered surfaces) on measurements at nanometric
separations. In this regime, roughness and gap distances are on the
same scale thereby obscuring the corresponding distance upon contact
between surfaces (or absolute measurements of the
separations)~\cite{DelRio05}. While predictions based on perturbative
approximations such as PFA played a key role in early
experiments~\cite{klimchistkaya96,genet03,maianeto05,pala,DeccaLo05},
the push to explore Casimir forces at shorter separations has
motivated more nuanced and precise theoretical studies of
roughness~\cite{Zwol08:rough,Broer12,Broer13}. For instance, recent
Casimir-force measurement in the sphere--plate geometry demonstrated
that at separations $\lesssim 100$nm, roughness effects manifest as
strong deviations in the power-law scaling of the force, leading to
large discrepancies $\sim 100\%$ between exact and perturbative
theoretical predictions~\cite{Zwol08:rough}. Application of
brute-force techniques~\cite{RodriguezMc09:PRA,McCauleyRo10:PRA} has
made it possible to perform reliable statistical studies of roughness
effects on the force at separations comparable to the root-mean square
of the height fluctuations, where the impact of shape, size, and
contribution of spurious peaks or asperities is
non-negligible~\cite{Broer12}.

The failure of PFA is even more pronounced in complicated, multi-body
geometries, such as the structure shown in
Fig.~\ref{fig:cas-collage}(e), involving two metallic co-planar
waveguides suspended above adjacent metal plates.  Using brute-force
numerics, Ref.~\cite{Rodriguez07:PRL} found that the attractive
Casimir force per unit length between the waveguides varies
non-monotonically as a function of their separation from the plates,
decreasing and then increasing as the plate separation decreases from
infinity to zero. It was found that the force deviates from PFA by
approximately $40\%$ in the limit of two isolated waveguides
($d\to\infty$) and is largest when the adjacent plates come into
contact with the waveguides, in which case the structure approaches
the Casimir ``piston-like'' geometry studied previously by analytical
means~\cite{Cavalcanti04}. More interestingly, application of simple
pairwise, line-of-sight approximations like PFA in which one sums the
lateral-force contribution of the plates to the waveguides, predicts a
monotonically increasing force with increasing plate
separation~\cite{Zaheer07}. Extensions of this structure to
cylindrical waveguides~\cite{Rahi07,Rodriguez-Lopez09} and exotic
materials~\cite{McCauleyRo10:PRA} have been explored, along with
explanations of the non-monotonicity which arises from non-additive,
screening effects~\cite{Rahi07}.  A recent experiment by Chan
et. al. measured force gradients in a related geometry via the novel,
integrated setup described in Section~\ref{sec:cas-exps}, involving a
silicon beam and an electrode suspended above a silicon plate, whose
cross-section is shown
Fig.~\ref{fig:casgeom}(c)~\cite{Zou13}. Although a significant step in
the right direction, the experiment was performed at lengthscales and
separations that precluded significant multi-body effects from being
observed.

Aside from non-monotonicity, one might ask whether it is possible to
significantly reduce or even flip the sign of the force from
attractive to repulsive. Design principles for achieving repulsion
have eluded theorists and experimentalists for decades, and currently
there are only a handful of ways of doing so. Materials approaches
date back to the early days of Lifshitz's theory and commonly involve
fluids~\cite{Dzyaloshinskii61,isra,Parsegian06,RodriguezMc10,RodriguezMc10,RodriguezWo10,Reid11,RodriguezReid13}
and/or exotic
materials~\cite{Boyer74,Leonhardt07,Rosa08:PRL,Zhao09,RodriguezLopez14}. In
fact, the influence of repulsive Casimir forces on fluid suspensions
and on wetting phenomena in thin films~\cite{Lamoreaux07:phystoday}
motivated much of the foundational work of Lifshitz,
Dzyaloshinski{\u{\i}}, and Pitaevski{\u{\i}}~\cite{Dzyaloshinskii61},
and has since played a crucial role in colloidal physics and
microfluidics~\cite{isra,Lamoreaux07:phystoday,Stone01}. Repulsion can
arise for instance in situations involving dissimilar bodies immersed
in fluids, a phenomenon that was first predicted in the van der Waals
regime of near-field dipolar interactions. Specifically, the van der
Waals force between two fluctuating dipoles of different
polarizabilities can flip sign, becoming repulsive when immersed in a
fluid of intermediate polarizability. Generalizations of this fluid
effect to situations involving macroscopic bodies and retardation were
performed in the 1960s~\cite{Dzyaloshinskii61} but only recently
observed in experiments~\cite{Munday09}. As illustrated in
Fig.~\ref{fig:casmat}(c), measurements by Munday et. al. reveal that
the long-range Casimir force between a gold-coated sphere and a
silicon plate immersed in bromobenzene is repulsive; in contrast, the
force becomes attractive when the plate is replaced by a thick gold
film. Because such repulsions depend on the dielectric properties of
materials over a wide range of wavelengths, not all fluid--material
combinations lead to repulsion. Current experimental efforts are
shedding light on these issues by developing tighter bounds on
material properties of common fluids and by exploring broader sets of
material configurations~\cite{Zwol09,Zwol10}. Recent theoretical works
have also focused on ways of exploiting material dispersion in
addition to nanostructuring to obtain stable fluid suspensions and
large temperature
corrections~\cite{Rodriguez08:PRL,RodriguezMc10:PRL,RodriguezWo10,McCauleyRo11:fluid,Phan12,Maofeng14}.

It has proven significantly more challenging to procure unusual
effects in vacuum-separated structures, especially in the original
geometry of two semi-infinite half-spaces where the only degrees of
freedom are the choice of materials. Early attempts to obtain
repulsion focused primarily on exotic materials such as perfect
magnetic conductors~\cite{Boyer74}, and more recently gain
media~\cite{Leonhardt07} and Chern
insulators~\cite{RodriguezLopez14}~\footnote{Application of Lifshitz
  theory to non-passive materials like gain media require care, since
  such media violate fundamental assumptions of thermodynamic
  equilibrium~\cite{Raabe07}.}. One trend in the past few years has
been to study ``metamaterials'' formed from metals and dielectrics
arranged into complex microstructures with sub-wavelength (nanometric)
features that lead to exotic effects at infrared wavelengths. Examples
include effective magnetodielectric, chiral and paramagnetic
metamaterials that were initially predicted to strongly alter Casimir
interactions~\cite{Henkel05,Rosa08:PRL,Zhao09,Ma14}. Unfortunately,
while metamaterials have proven useful in classical-force designs (as
described above), they often rely on narrow-band resonances that turn
out to have a negligible effect on the force once the full spectrum
and realistic microstructures are
considered~\cite{Rosa09,Silveirinha10,McCauleyZh10}. In fact, general
arguments based on the passivity of dielectric (non-magnetic)
materials discount the possibility of repulsion in general multilayer
systems~\cite{Lambrecht97}. There is also growing interest in
understanding the influence of magnetism on the force in realistic
materials. For instance, Ref.~\cite{Klimchitskaya10:magnetics}
recently showed that magnetic effects in common diamagnetic and
paramagnetic materials have little to no impact on the force, in
contrast to ferromagnetic materials such as Co and Fe that can
potentially enable temperature-tunable interactions if one operates
near their phase transition
temperature~\cite{Klimchitskaya10:magnetics}. Follow-up experiments
with Ni-coated surfaces operating near room temperature have
demonstrated small changes to the force~\cite{Banishev13:ni}.
Finally, reductions in the force can also be achieved by modifying
material properties at the atomic and electronic level, such as by
varying the doping~\cite{Duraffourg06,LambrechtPi06} or carrier
density~\cite{Chen06:mems} of semiconductors like silicon, or by
employing aereogels with varying porosity~\cite{EsquivelSirvent07}.

Structural approaches have only recently begun to be explored, thanks
largely to the emergence of powerful theoretical techniques. These
sophisticated formulations have also been used to establish general
limits on the kinds of repulsive interactions that can arise through
structural considerations alone. For instance, one can show that the
force between multilayer films~\cite{Lambrecht97} or mirror-symmetric
objects~\cite{KennethKl06} separated by vacuum is always attractive,
and also that stable suspensions (local equilibria) between
vacuum-separated metal/dielectric objects are
impossible~\cite{Rahi10:PRL}. There remain however many other
circumstances of interest. In planar systems the only structural
degrees of freedom are the thicknesses of films, which can be varied
to obtain non-negligible reductions in the
force~\cite{Duraffourg06,Palasantzas08}. For instance, decreasing the
thickness of a metal film beyond its corresponding bulk skin-depth in
the infrared to ultraviolet regime, i.e. $\lesssim 10$nm for common
metals, causes the film to become increasingly transparent
(non-reflective) and this results in a corresponding decrease in the
Casimir force. The first observations of the skin-depth effects were
performed by Lisanti et. al. in 2005 by employing an AFM setup
involving a metal-coated polystyrene sphere positioned above a
micromachined torsional oscillator~\cite{lisanti,Iannuzzi06}. A
reduction in the Casimir force on the order of 30\% was obtained with
thin $\sim 0.1$nm films, in agreement with theoretical predictions
based on the Lifshitz theory. Another experiment
[Fig.~\ref{fig:casmat}(b)] measuring forces between a $100\mu$m
gold-coated sphere and gold-coated silica plate in the 50--150nm
separation range demonstrated a 50\% reduction in the Casimir force
when switching from gold to indium tin oxides (ITO) coatings due to
the latter's significantly smaller plasma
frequency~\cite{Man09}. Other efforts include investigations of forces
between graphene sheets~\cite{GomezSantos09}, whose unique mechanical,
electrical, and optical properties are proving useful in a wide range
of applications. For instance, the experiment of Banishev
et. al. recently demonstrated [Fig.~\ref{fig:casmat}(a)] that
depositing a graphene sheet on a silicon/silica substrate can
increases the force by $\sim 10\%$~\cite{Banishev13:graphene}.

Predictions of repulsion in highly non-planar systems have also
emerged in the past few years. For instance, two interleaved objects
such as the zipper-like glide-symmetric structure shown in
Fig.~\ref{fig:cas-collage}(f) involving interleaved metal brackets
attached to parallel plates, can repel one another due to a
competition between lateral and normal forces. In particular,
depending on the separation, the attractive lateral force between the
brackets can dominate the interaction and either contribute toward or
act against the attractive plate--plate and plate--bracket
interactions, the net result of which is attraction at large and short
separations and repulsion at intermediate separations. Although such
an effect can be qualitatively understood from pairwise interactions,
it was recently confirmed by exact calculations that also illustrate
the dramatic, quantitative failure of PFA in highly non-planar
structures~\cite{RodriguezJo08:PRA}. Experimental demonstrations of
this effect are currently being pursued. A more direct and less
ambiguous prediction of geometry-induced repulsion (in non-interleaved
geometries) was recently made by Levin et. al., demonstrating that the
force on a small elongated object centered above a plate with a hole
switches from attractive to repulsive as the object approaches the
plate [Fig.~\ref{fig:cas-collage}(g)]. This effect was explained via a
symmetry argument in an idealized geometry involving an infinitesimal
particle or dipole above an infinitesimally thin perfect-metal plate
with a hole, and validated via brute-force calculations in realistic
geometries and materials. Although this was the first demonstration of
a geometry-based repulsive effect that cannot arise from pairwise
attractions, the magnitude of the force is unfortunately too small for
current experimental detection and other alternatives are currently
being explored~\cite{McCauleyRo11,Milton11}.

Finally, in addition to normal and lateral forces, quantum
fluctuations can also induced torques on objects that are free to
rotate. For example, torques can arise in anisotropic or birefringent
materials~\cite{Parsegian72,Barash78,Enk95:torque,Munday05,Romanowsky08}
or through
geometry~\cite{Razmi05,Kenneth00:torque,Rodrigues06:torque,MiltonPa10}. To
date, however, there have been no experimental observations of such
QED torques.

\subsection{Technological Impact}
\label{sec:cas-tech}

\begin{figure}[tb!]
  \centering
   \includegraphics[scale=0.47]{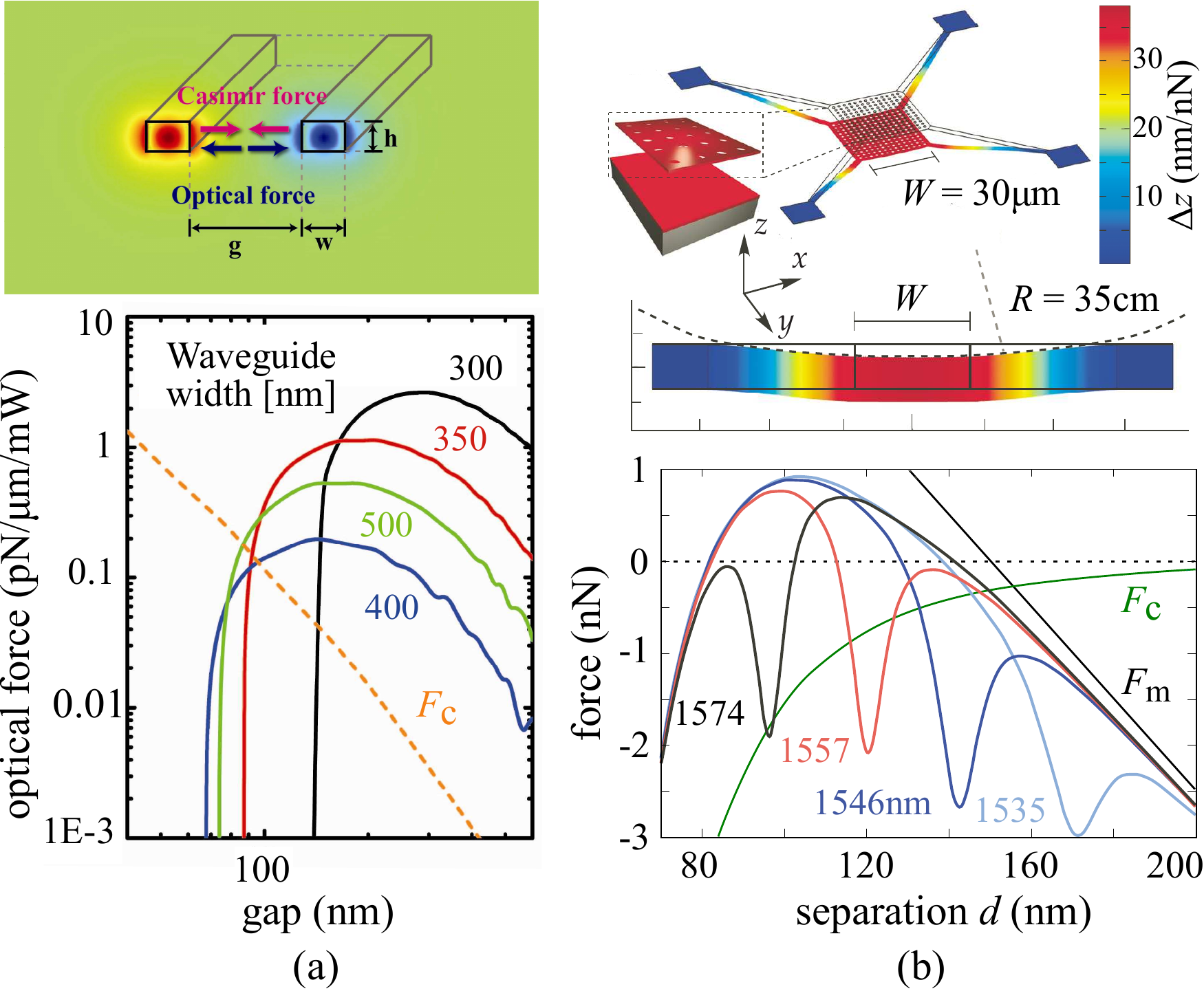}
   \caption{Theoretical investigations of the interplay between
     Casimir and optical gradient forces in two representative
     optomechanical structures (shown
     schematically)~\cite{Pernice10,RodriguezWo11}. (a) Optical (solid
     lines) and Casimir (dashed line) force per unit length and per mW
     of incident power between two rectangular silicon waveguides of
     height 220nm, plotted for multiple configurations of waveguide
     widths. Because the Casimir force increases more rapidly than the
     optical force, increasingly larger powers are required to
     counteract stiction in this system. (b) Total force (solid lines)
     on a $30\mu\mathrm{m}\times 30\mu\mathrm{m}$ PhC square membrane
     of 130nm thickness suspended above a silicon-on-insulator
     substrate by four deformable arms, as a function their
     surface--surface separation $d$.  The total force includes
     mechanical (dashed black), Casimir (dashed green) and
     anti-bonding gradient forces. The optical force is induced by
     normally incident light from above with power $P = 10$mW and
     (variable) wavelength $\lambda$. The stable position of the
     membrane varies depending on the incident wavelength, displaying
     bistability as $\lambda$ is swept from smaller to larger
     values. The schematic shows that most of the deformation occurs
     within the $L=30\mu$m arms supporting the membrane.}
   \label{fig:cas-optomech}
\end{figure}

The unavoidable impact of Casimir forces on MEMS has been a driving
force behind much of the recent progress discussed above, especially
as scaling issues and additional complexity continue to push devices
toward increasingly smaller (nanometric) scales. The adhesion or
stiction of neighboring movable parts in micromachines from
electrostatic or Casimir forces is of special concern, as described in
a number of seminal works~\cite{DelRio05,Serry98,Buks01:mems}. On the
other hand, the ability to tune Casimir forces in MEMS can potentially
lead to novel functionalities and architectures. Examples include
driven nonlinear electromechanical
oscillators~\cite{Serry95,hochan1,hochan2}, rack pinions based on
lateral forces~\cite{Ashourvan07}, and ultra-sensitive sensors based
on parametric amplifiers~\cite{Imboden14}. Recent theoretical
predictions even suggest that roughness can have a potentially
positive influence on the actuation dynamics of MEMS at nanometric
scales~\cite{Broer13}. The experiment by Chan et. al. in 2001
demonstrated that at short separations, the Casimir interaction
between a metal sphere attached to a movable oscillator and a plate
adds a nonlinear term to the otherwise linear mechanical restoring
force on the sphere, resulting in a number of interesting nonlinear
dynamical effects such as bistability and
hysteresis~\cite{hochan1,hochan2}. Meanwhile, experiments such as
those by Zhou et. al. demonstrating in-situ measurements of Casimir
forces in on-chip integrated silicon devices continue to bring Casimir
forces into the forefront of MEMS
technology~\cite{Zou13}. Anti-stiction mechanisms based on lateral
forces~\cite{RodriguezJo08:PRA} in similar integrated platforms are
currently being explored.

The integration of Casimir forces and optomechanical devices offers
even more degrees of freedom for achieving designable actuation
effects in new generations of NOEMS. For instance, repulsive optical
forces can be used to combat stiction, bringing to bear the unique
competitive edge of gradient-force designs in optomechanical
systems~\cite{Pernice10,RodriguezWo11}. The interplay and expression
of Casimir and gradient forces were recently studied in systems based
on coupled waveguides~\cite{Pernice10} and PhC
membranes~\cite{RodriguezWo11}. Simulations reveal
[Fig.~\ref{fig:cas-optomech}] that all-optical, anti-bonding forces
excited by incident light at relatively low powers ($\lesssim
10~\mathrm{mW}$) can be used to control the onset of stiction at small
separations ($\lesssim 100$nm), and can also greatly influence the
static and dynamical mechanical state of these
systems~\cite{RodriguezWo11}. The large interaction areas of PhC
membranes not only allow strong Casimir effects to arise at larger
separations, relaxing the signal-to-noise requirements for accurate
detection, but also provide a mechanism for the simultaneous actuation
and detection of Casimir forces. Initial experiments in a related
structure [Fig.~\ref{fig:dSOI}(b)] demonstrate nanometer-range pulling
($\sim 1$nm$/1$mW) and optomechanically induced optically bistability
effects~\cite{Woolf13, Hui13}. By combining the above ingredients, the
prospect of real-time monitoring and control of MEMS/NEMS through
optomechanical degrees of freedom is becoming increasingly possible.

\section{Concluding Remarks}

Because of the infinite variety of possible geometries, and the highly
nonlinear dependence of electromagnetic fields and forces on the
geometric and material parameters, it is impossible to be certain what
the future holds for optomechanical interactions in
nanophotonics. This is, perhaps, especially true for
fluctuation-induced interactions, where both theory and experiment have
only just begun to move past the simplest planar structures of the
past. The combination of optomechanical interactions with other
phenomena, such as nonlinear optics (including nonlinear-gain media
and lasers), exotic plasmonic metamaterials, or fluid-mechanical
effects, has been only lightly explored at best.  One possible avenue
for such explorations, which has been gaining ground in other areas of
photonic design but is thus far little-used in the regime of classical
and quantum forces, is computational ``inverse design'' in which
hundreds or thousands of parameters are explored systematically by
optimization techniques in order to discover new geometries and new
regimes of operation~\cite{Bendsoe03}. 


\end{document}